\documentclass[tradiabstract]{aa} % for the abstract without structuration 
\newcommand{\hajduk}{HZG08}
\newcommand{\hajduks}{HZG08\ }
\newcommand{\cpd}{CPD$-$56$^\circ$8032}
\newcommand{\cpds}{CPD$-$56$^\circ$8032\ }
\usepackage{graphicx}
\usepackage{txfonts}
\begin{document}
\title{The influence of binarity on dust obscuration events in the planetary nebula M~2-29 and its analogues\thanks{Based on observations made with the Very Large Telescope at Paranal Observatory under program ID 079.D-0764(A).}
}
   \author{B. Miszalski
   \inst{1,2,3}
           \and
           J. Miko{\l}ajewska
           \inst{4}
           \and
           J. K\"oppen
           \inst{2,5,6}
           \and
           T. Rauch   
           \inst{7}
           \and
           A. Acker
           \inst{2}
           \and
           M. Cohen
           \inst{8}
           \and\\
           D. J. Frew
           \inst{3}
           \and
           A. F. J. Moffat
           \inst{9}
           \and
           Q. A. Parker
           \inst{3,10}
           \and
           A. F. Jones
           \inst{11}
           \and
           A. Udalski
           \inst{12}
          }

\institute{
Centre for Astrophysics Research, STRI, University of Hertfordshire, College Lane Campus, Hatfield AL10 9AB, United Kingdom\\
\email{b.miszalski@herts.ac.uk}
\and
Observatoire Astronomique, Universit\'e de Strasbourg, 11 rue de l'Universit\'e, F-67000, Strasbourg, France
\and
Department of Physics, Macquarie University, Sydney, NSW 2109, Australia
\and
Nicolaus Copernicus Astronomical Centre, Bartycka 18, PL-00716, Warsaw, Poland\\
\email{mikolaj@camk.edu.pl}
         \and
         International Space University, Parc d'Innovation, 1 Rue Jean-Dominique Cassini, F-67400, Illkirch-Graffenstaden, France
         \and
         Institut f\"ur Theorestische Physik und Astrophysik, Universit\"at Kiel, D-24098, Kiel, Germany
         \and
         Institute for Astronomy and Astrophysics, Kepler Center for Astro and Particle Physics, Eberhard Karls University, Sand 1, D-72076, T\"ubingen, Germany
         \and
         Radio Astronomy Laboratory, University of California, Berkeley, CA 94720, USA
         \and
         D\'ept. de physique, Univ. de Montr\'eal C.P. 6128, Succ. Centre-Ville, Montr\'eal, QC H3C 3J7, and Centre de recherche\\ en astrophysique de Qu\'ebec, Canada
         \and
         Australian Astronomical Observatory, Epping, NSW 1710, Australia
         \and 
         School of Chemical and Physical Sciences, Victoria University of Wellington, PO Box 600, Wellington, New Zealand
         \and
         Warsaw University Observatory, Al. Ujazdowskie 4, PL-00-478, Warsaw, Poland
         }
   \date{Received -; accepted -}

  \abstract
  {The central star of the planetary nebula (CSPN) M~2-29 shows an extraordinary R Coronae Borealis-like fading event in its optical lightcurve. The only other CSPN to show these events are \cpds (Hen~3-1333) and V651 Mon (NGC~2346). 
  Dust cloud formation in the line of sight appears responsible but the exact triggering mechanism is not well understood. 
  Understanding how planetary nebulae (PNe) trigger dust obscuration events may help understand the same process in a wide range of objects including Population-I WC9 stars, symbiotic stars and perhaps Asymptotic Giant Branch (AGB) stars with long secondary periods (LSPs). A binary scenario involving an eccentric, wide companion that triggers dust formation via interaction at periastron is a potential explanation that has been suggested for LSP variables. Model fits to the lightcurves of \cpds and M~2-29 show the dust forms in excess of 70 AU at the inner edge of a dust disk. In the case of \cpds this radius is far too large to coincide with a binary companion trigger, although a binary may have been responsible for the formation of the dust disk. We find no direct evidence to support previous claims of binarity in M~2-29 either from the OGLE lightcurve or deep medium-resolution VLT FLAMES spectroscopy of the CSPN. We classify the CSPN as Of(H) with $T_\mathrm{eff}=50\pm10$ kK and log $g=4.0\pm0.3$. We find a mean distance of $7.4\pm1.8$ kpc to M~2-29 at which the $M_V=-0.9$ mag CSPN could potentially hide a subgiant luminosity or fainter companion. A companion would help explain the multiple similarities with D'-type symbiotic stars whose outer nebulae are thought to be bona-fide PNe. The 7.4 kpc distance, oxygen abundance of 8.3 dex and Galactic coordinates ($\ell=4.0$, $b=-3.0$) prove that M~2-29 is a Galactic Bulge PN and not a Halo PN as commonly misconceived. 
   }

   \maketitle
\section{Introduction}
Long-term photometric monitoring surveys such as MACHO (Alcock et al. 1997) and OGLE (Udalski 2009) have opened a remarkable window into the decadal variability of rare, short-lived phases of stellar evolution. Perhaps least understood are those variables that exhibit fading events due to dust obscuration. These include R Coronae Borealis stars (RCB, Clayton 1996), Population-I WC9 stars (Veen et al. 1998; Kato et al. 2002), symbiotic stars (Miko{\l}ajewska et al. 1999; Mennickent et al. 2008; Gromadzki et al. 2009), PNe (M\'endez, Gathier \& Niemela 1982; Cohen et al. 2002; Hajduk et al. 2008, hereafter \hajduk) and perhaps AGB stars with long secondary periods or LSPs (Wood et al. 1999, 2004; Soszy{\'n}ski 2007; Nicholls et al. 2009; Wood \& Nicholls 2009). 

In all these systems but the LSP variables, it is reasonably well established that ongoing dust cloud formation in the line of sight is responsible for the fading events. The exact mechanism that triggers the dust formation is not always known but shock waves are believed to play an important role (Fleischer et al. 1992; Goeres \& Sedlmayr 1992, hereafter GS92; Veen et al. 1998). Stellar pulsations or interactions with a binary companion can help produce these shock waves. 
There are no binary RCB stars known which leaves pulsations as the only possible factor (e.g. Lawson et al. 1990). Amongst Population-I WC stars the non-WC9 types form amorphous carbon dust via wind-wind collision (WWC) in binary systems (Tuthill, Monnier \& Danchi 1999; Williams 2008). The WC9 stars appear to be different in that while the majority do form C-based dust, it's unclear whether this is due to WWC or shocks associated with the clumpy winds of single stars. Ground-based observations of WC9 stars show that most do not vary above $\sigma$=5--10 mmag (Fahed, Moffat \& Bonanos 2009) which strongly suggests pulsations are not responsible for dust formation in these stars. 
Symbiotic stars are by definition binaries and pulsations are common in the evolved components (e.g. Miko{\l}ajewska 2001). Pulsations are known to occur mostly in H-deficient PNe (e.g. Ciardullo \& Bond 1996; Gonz\'alez P\'erez, Solheim \& Kamben 2006), but our knowledge of binarity in PNe is still in its infancy especially for orbital periods $\ga$1 day (Miszalski et al. 2009, 2010b; De Marco 2009).

The origin of LSPs in AGB stars has a long history of debate but no satisfactory explanation for them exists (Nicholls et al. 2009). Dust is clearly associated with the phenomenon but the presence of dust alone cannot distinguish between competing scenarios (Wood \& Nicholls 2009). An alternative way to view the problem is to examine PNe as the progeny of AGB stars with LSPs and to look for evidence that would support a given scenario. Of all the candidate scenarios a wide binary model (e.g. Wood et al. 1999) would be the most straightforward to verify with the safe assumption that the companion survives the PN phase. In this scenario a very low mass companion in a large eccentric orbit interacts at periastron triggering dust formation. The dust formed may remain clumpy or it could settle into a stable disk structure. If a disk is present prior to PN ejection the dust geometry would conspire to produce a bipolar or at least asymmetric nebula (Bond \& Livio 1990), while the disk itself would not necessarily survive the ejection event. This relationship with asymmetric PNe (see also Wood et al. 2004), and particularly the strong resemblance between fading events in PNe and LSPs, makes PNe a potentially valuable tool to address the longstanding LSP problem. 

\textbf{M~2-29 (PN G004.0$-$03.0)} is the least studied amongst the three PNe known to show dust obscuration events, partly because this behaviour was only discovered recently (\hajduk). M~2-29 is infamous for its apparently low oxygen abundance of 7.4$\pm$0.1 dex which lies in the metal poor regime of Galactic Halo PNe (Pe\~na, Torres-Peimbert \& Ruiz 1991, 1992, hereafter P91, P92; Howard et al. 1997). Torres-Peimbert et al. (1997, hereafter TP97) found M~2-29 to have a high density ($n_e\ge10^6$ cm$^{-3}$) circumstellar nebula in \emph{HST} observations and consequently remarked upon the spurious nature of ground based abundance measurements. We obtained medium resolution integral field spectroscopy of M~2-29 with the VLT to measure its chemical abundances prior to the discovery of the fading event. Miszalski et al. (2009) confirmed the dense circumstellar nebula of TP97 and further suggested that it may indicate the nucleus is a symbiotic star. Here we expand upon this claim within the context of dust obscuration events in PNe. 

This paper is structured as follows. Section \ref{sec:known} reviews dust obscuration in V651 Mon and \cpd. Section \ref{sec:lc} discusses the photometric variations of M~2-29 with a particular focus on whether there is any evidence for binarity. Section \ref{sec:obs} presents the first deep, medium-resolution, spatially-resolved optical spectroscopy of M~2-29. Classification of the CSPN of M~2-29 is discussed in Sect. \ref{sec:cspn} and the properties of its nebula are covered in Sect. \ref{sec:neb}. The dust properties of M~2-29 are discussed in Sect. \ref{sec:dust}. Section \ref{sec:sy} outlines the similarities between M~2-29 and symbiotic stars, and we conclude in Sect. \ref{sec:conclusion}.

\section{Dust obscuration in NGC~2346 and Hen~3-1333}
\label{sec:known}
Apart from M~2-29 only V651 Mon of NGC~2346 (PN G215.6$+$03.6; M\'endez et al. 1982) and the [WC10] nucleus \cpds of Hen~3-1333 (PN G332.9$-$09.9; Pollacco et al. 1992; Jones et al. 1999; Cohen et al. 2002) were known to show dust obscuration events. If there is a common binary explanation for the fading events in AGB stars with LSPs and these PNe, then it is fundamental to establish the binary status or otherwise of these PNe. A key question is whether clearly defined or isolated fading events repeat periodically presumably due to the influence of a companion. An additional tool that may be used to infer binarity is the detection of a (circumbinary) dust disk with Keplerian kinematics (Van Winckel 2003; Bujarrabal et al. 2005; Deroo et al. 2007). Binarity appears to be a prerequisite for disk formation but much further observational evidence is required to test this hypothesis (Van Winckel et al. 2009; Chesneau 2010). 

\textbf{V651 Mon} consists of an A5V star and an invisible hot sub-dwarf in a 16 day period seen in both photometric and radial velocity observations (M\'endez \& Niemela 1981; M\'endez et al. 1982). Undoubtedly the longest monitored CSPN, V651 Mon has been observed since the 1970s (see e.g. Schaefer 1983, 1985; Kohoutek 1995) and is also monitored by the American Association of Variable Star Observers (AAVSO\footnote{http://www.aavso.org}) and the All Sky Automated Survey (ASAS, Pojmanski 2002). Isolated minima as well studied as the one observed by Kato et al. (2001) are difficult to glean from the early literature, while the early AAVSO data show what appears to be a prolonged minimum up until HJD$-2450000=-3900$ days. This could correspond to the quasi-stationary Phase C described by GS92 where an equilibrium persists between dust formation and radiative pressure. Table \ref{tab:minima} presents the epochs and duration of recent isolated minima from Kato et al. (2001) and ASAS. It is inappropriate to claim any periodicity until further observations of minima are obtained. If a period of $\sim$5 years is adopted, then a minimum should have been observed around HJD$-2450000\sim1800$ days, but it is possible the conditions for such an event were not suitable for grain growth to occur.

If fading events in V651 Mon were to be triggered by a wide binary it is clearly not either component of the observed 16 day binary. Unfortunately the radial velocities presented in the literature are too few to confirm or deny such a companion. A circumbinary disk seems unlikely to surround V651 Mon since the A5V component of the binary CSPN is only slightly reddened outside fading events (M\'endez \& Niemela 1981; M\'endez et al. 1985). The evacuated inner zone or `void' (Balick 2000) also means the survival of a disk after the PN phase is unlikely. Nevertheless, high-resolution (interferometric) observations of NGC~2346 are urged. 

\begin{figure}
   \begin{center}
      \includegraphics[scale=0.35,angle=270]{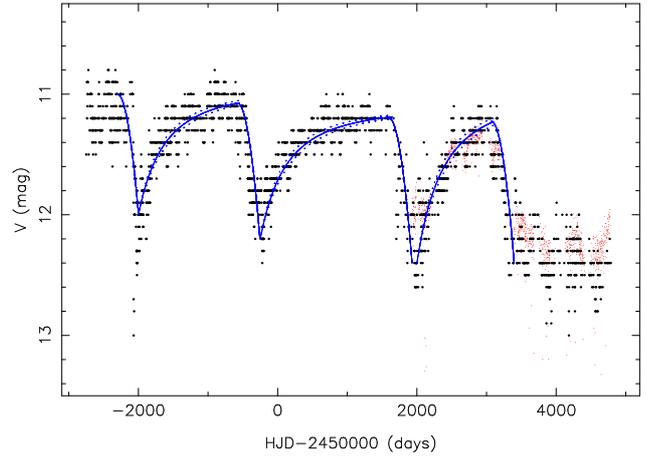}
   \end{center}
   \caption{An updated version of the \cpds lightcurve published by Cohen et al. (2002), black points, and ASAS (Pojmanski 2002), red points. GS92 model fits for $R_d=97\pm11$ AU and $v_\infty=225$ km s$^{-1}$ (blue lines) to the lightcurve have (left to right) min/max magnitudes of 12.0/11.0, 12.2/11.1 and 12.4/11.0, and last 1450, 1900 and 1100 days.}
   \label{fig:cpd}
\end{figure}

\textbf{\cpds} is the cool [WC10] nucleus of Hen~3-1333 which De Marco \& Crowther (1998) found to have $T_\mathrm{eff}=30$ kK, $v_\infty=225$ km s$^{-1}$ and a high mass-loss rate of $\dot{M}=4\times10^{-6}$ $M_\odot$ yr$^{-1}$. A dust disk of inner radius 97$\pm$11 AU was resolved by the VLTI (De Marco et al. 2002b; Chesneau et al. 2006) and dual-dust chemistry is present (Cohen et al. 2002). Figure \ref{fig:cpd} shows the visual lightcurve of \cpds observed by one of us (AFJ) that updates the previous version published by Cohen et al. (2002). The minima are tabulated in Tab. \ref{tab:minima} and, as for V651 Mon, we suspect \cpds is in a prolonged quasi-stationary state of minimum after HJD$-2450000=3500$ days. 

Cohen et al. (2002) described a number of plausible scenarios to describe the minima which have largely remained untested. Phenomenologically the lightcurve of \cpds resembles the events in RCB stars but over much longer timescales (e.g. Alcock et al. 2001; Tisserand et al. 2008). To gain further insight into the problem we can model the lightcurve using either the RCB star model of GS92 or the WC9 model of Veen et al. (1998). The main free parameters in these models are the radius $R_d$ between the star and the forming dust cloud, and the dust outflow velocity i.e. the terminal velocity of the stellar wind $v_\infty$. The WC9 model of Veen et al. (1998) did not match the curvature of the ingress for any $R_d$ and was not considered further. For the ingress we used Eqn. 33a of GS92 which is a simple quadratic independent of both $v_\infty$ and $R_d$ that only describes dust grain growth. For the egress we used Eqn. 35 of GS92:
\begin{equation}
%\[
   m(t) - m(t_1) = \left( m(t_0)-m(t_1) \right)\frac{R_d^2}{(R_d+v_\infty t)^2} 
   \label{eqn:dust}
%   \]
\end{equation}
where $m(t_0)$ and $m(t_1)$ are the magnitudes at light minimum and maximum, respectively. A remarkably good fit was found using the empirically determined $R_d=97\pm11$ AU and $v_\infty=225$ km s$^{-1}$ (Fig. \ref{fig:cpd}). 

The quality of the fit suggests the same physics apply to fading events in PNe as RCB stars despite the very different time scales involved. \emph{In other words, it is likely that the dust forms at the inner edge of the dust disk in \cpd}. As $R_d$ is 3--4 orders of magnitude larger than in RCB stars, presumably because of the stronger ionising field of the CSPN, the wind densities would be insufficient to support dust formation (De Marco et al. 2002a). The only way to reach the high densities required would be either with a dust disk or via wind-wind interaction with a companion. The dust disk is already established and the latter scenario is unlikely in \cpds because for an assumed orbital period of $\sim$5 years, a primary mass of 0.6 $M_\odot$ and a main-sequence companion of 0.4 $M_\odot$, the orbital separation, and therefore $R_d$, would be at most $13\pm2$ AU. This radius is far too small to explain the fading events in \cpds.

In summary, there is insufficient evidence for obscuration events in PNe to be triggered by a wide interacting companion. Firstly, there is a lack of a strong periodic repeatability from both V651~Mon and \cpds, and secondly, we have excluded a companion to \cpds in a $P\sim5$ year orbit as the trigger is based on the substantially larger dust formation radius at $97\pm11$ AU. While binarity may be responsible for the formation of a dust disk, binarity itself does not seem to play a strong role in the dust formation events. All the above suggests the events are triggered by stochastic instabilities in the dust disk itself, perhaps as a result of clumpy, high mass-loss rate winds from the central stars. Further observations of V651 Mon to clearly establish the presence of a dust disk are urged.

%Events last $\sim$300--400 days for V651 Mon and $\sim$800--1000 days in \cpd.}
\begin{table}
   \caption{Isolated minima in V651 Mon and \cpd.}
   \label{tab:minima}
   \centering
   \begin{tabular}{rrrr}
      \hline\hline
       V651 Mon & $\Delta t$ & \cpd & $\Delta t$\\
       (HJD$-2450000$) & (years) & (HJD$-2450000$) & (years)\\
      \hline
      500 & 0 & -2000 & 0 \\    
      3250 & 7.5 & -250 & 4.8  \\
      4900 & 4.5 & 2000 & 6.2  \\ 
      -    & -   &  3500 & 4.1 \\
      \hline
   \end{tabular}
\end{table}

\section{The OGLE lightcurve of M~2-29}
\label{sec:lc}
\begin{figure}
   \begin{center}
      \includegraphics[angle=270,scale=0.35]{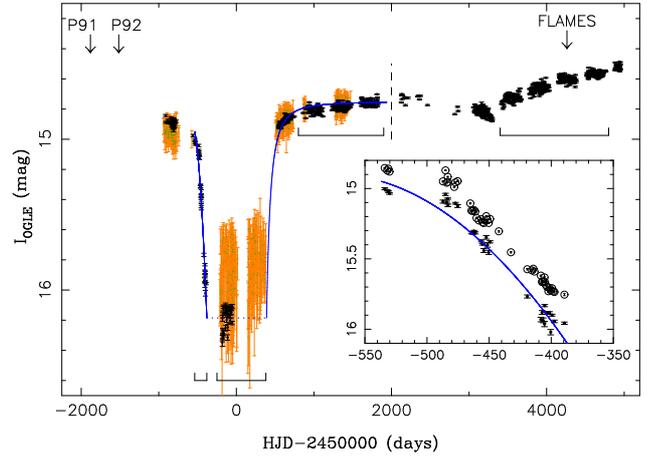}
   \end{center}
   \caption{OGLE $I$-band (black points) and MACHO Cousins $R$ (green points, shifted $-$0.9 mag) lightcurves of M~2-29. The blue lines are GS92 RCB star model fits. The inset shows the ingress in the old (circles) and final (points) OGLE-I lightcurves. Spectroscopic observations are arrowed and ranges analysed for periodic variability marked. The dashed vertical line demarcates OGLE-II from OGLE-III data.}
   \label{fig:lc}
\end{figure}

Figure \ref{fig:lc} depicts the OGLE $I$-band lightcurve after final processing to place all OGLE data on a uniform photometric scale (Udalski et al. 2008). The MACHO Cousins $R$ lightcurve is also shown to indicate the duration of the deep minimum (Alcock et al. 1997; Lutz et al. 2010).\footnote{The correct MACHO FTS identification number is 101.21568.184 and not 101.21568.639 as listed by Lutz et al. (2010).} In the following we examine the large-scale and short-scale $I$-band variations, as well as the OGLE $V$-band data, to search for evidence of binarity. 

\subsection{Large-scale variations}
\label{sec:large}
Zebrun (1998) first classified the CSPN of M~2-29 as an RCB star candidate, but it was not until \hajduks that the connection was made with the PN. As for \cpd, we found the fading event of M~2-29 could be successfully modelled using the simple RCB star model of GS92. Our model fit in Fig. \ref{fig:lc} using Eqn. \ref{eqn:dust} assumes an upper limit of $v_\infty=1200$ km s$^{-1}$ (Sect. \ref{sec:class}) and adopts $R_d=70\pm10$ AU, $I=16.18$ at light minimum and $I=14.75$ at maximum. The ingress model assumes total occultation of the star which may explain the difficulty in fitting the preceeding slow decline. 

The simple assumptions of the GS92 model are by Ockham's Razor preferable over more complicated alternative scenarios that involve occultation by a large dust disk (e.g. \hajduk). The larger $v_\infty$ creates a more rapid recovery than in \cpds and the very large dust formation radius strongly suggests a dust disk is present in M~2-29 (see also Sect. \ref{sec:dust}). The double-peaked [O~III] profile detected by Gesicki et al. (2010) may be evidence for Keplerian rotation in the dust disk following the example of EGB~6 (Su et al. 2010). 
An alternative view may be the profile reflects a small bipolar outflow emerging from the disk. 

The slow $\sim$0.4 mag variations at light maximum in M~2-29 are not seen in V651 Mon or \cpd. Only variations reflecting the 16 day orbital period are seen in V651 Mon. The timescales are again much greater than in RCB stars where they are pulsation-driven. As we find no evidence for pulsations (Sect. \ref{sec:small}), we speculate that they are the product of a continuously variable (perhaps clumpy) mass-loss rate. Further time-resolved spectroscopy of the CSPN to track changes in stellar parameters should clarify their origin. At this stage it is highly premature to assign any type of periodicity to the fading events given only one is observed. \hajduks claimed to interpret the `dip' around $\mathrm{HJD}-2450000=3200$ days (i.e. the third OGLE-III season) to represent a `secondary eclipse' from which they derived an orbital period of $\sim$18 years. There are a number of problems with this interpretation. Firstly, the nomenclature of stellar eclipses cannot be applied to fading events, and secondly, there is no evidence for any declining trend in the adjacent photometry near the gap in coverage that would suggest a fading event had occurred. The latter point is supported by the photometry of the third OGLE-III season which monotonically increases rather than decreases towards earlier dates.

\subsection{Small-scale variations}
\label{sec:small}
\begin{figure}
   \begin{center}
      \includegraphics[angle=270,scale=0.35]{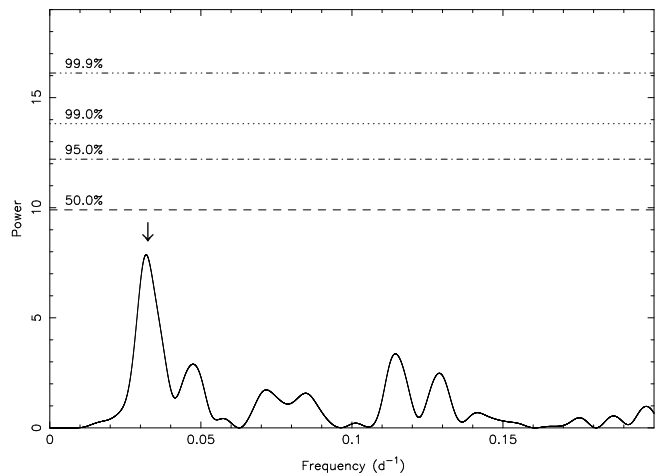}
   \end{center}
   \caption{Lomb periodogram of the detrended ingress into the fading event. Horizontal lines are the formal significance levels and the arrowed peak at 31.2 days is a likely sampling alias.}
   \label{fig:fourier}
\end{figure}

\begin{figure}
   \begin{center}
      \includegraphics[scale=0.275]{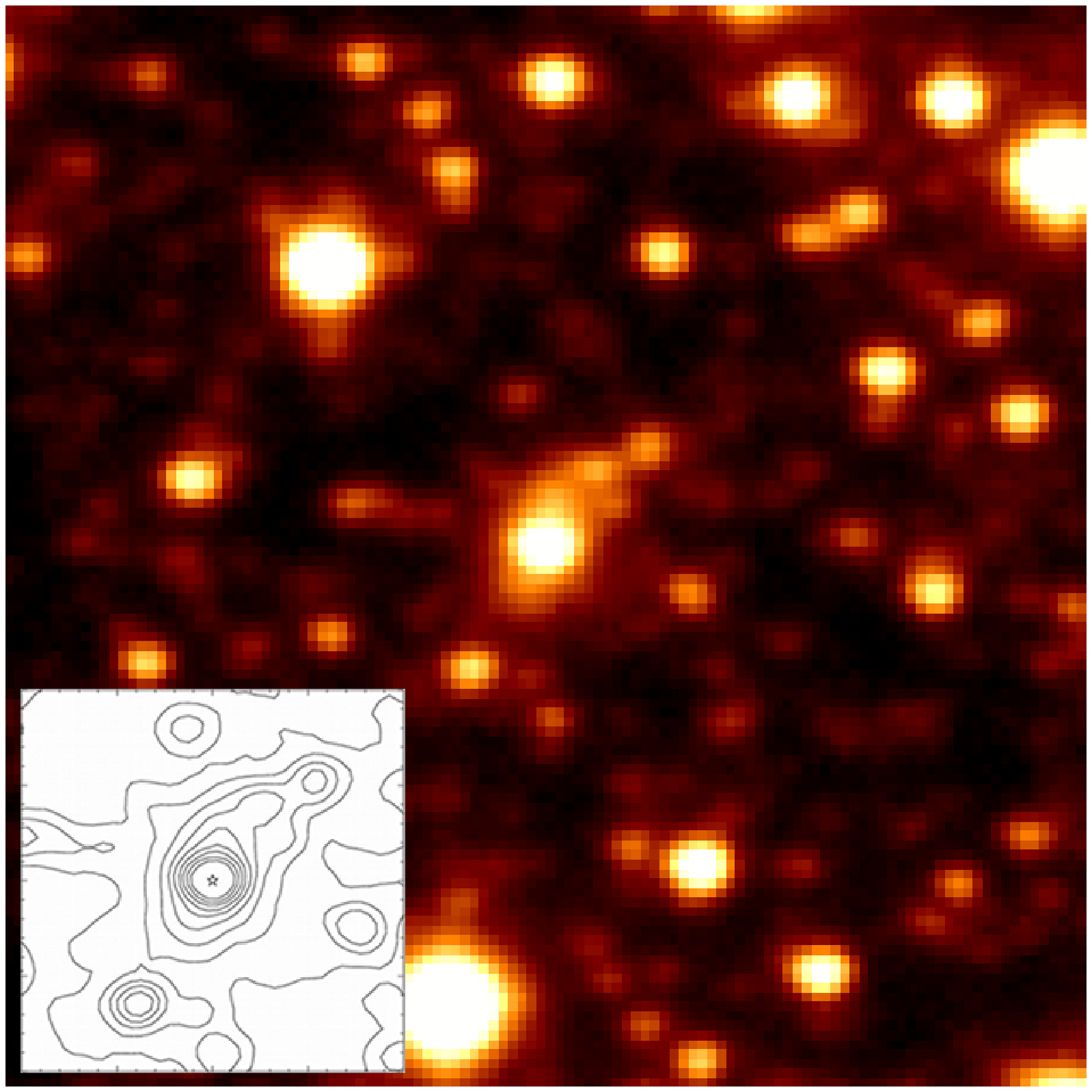}
      \includegraphics[scale=0.275]{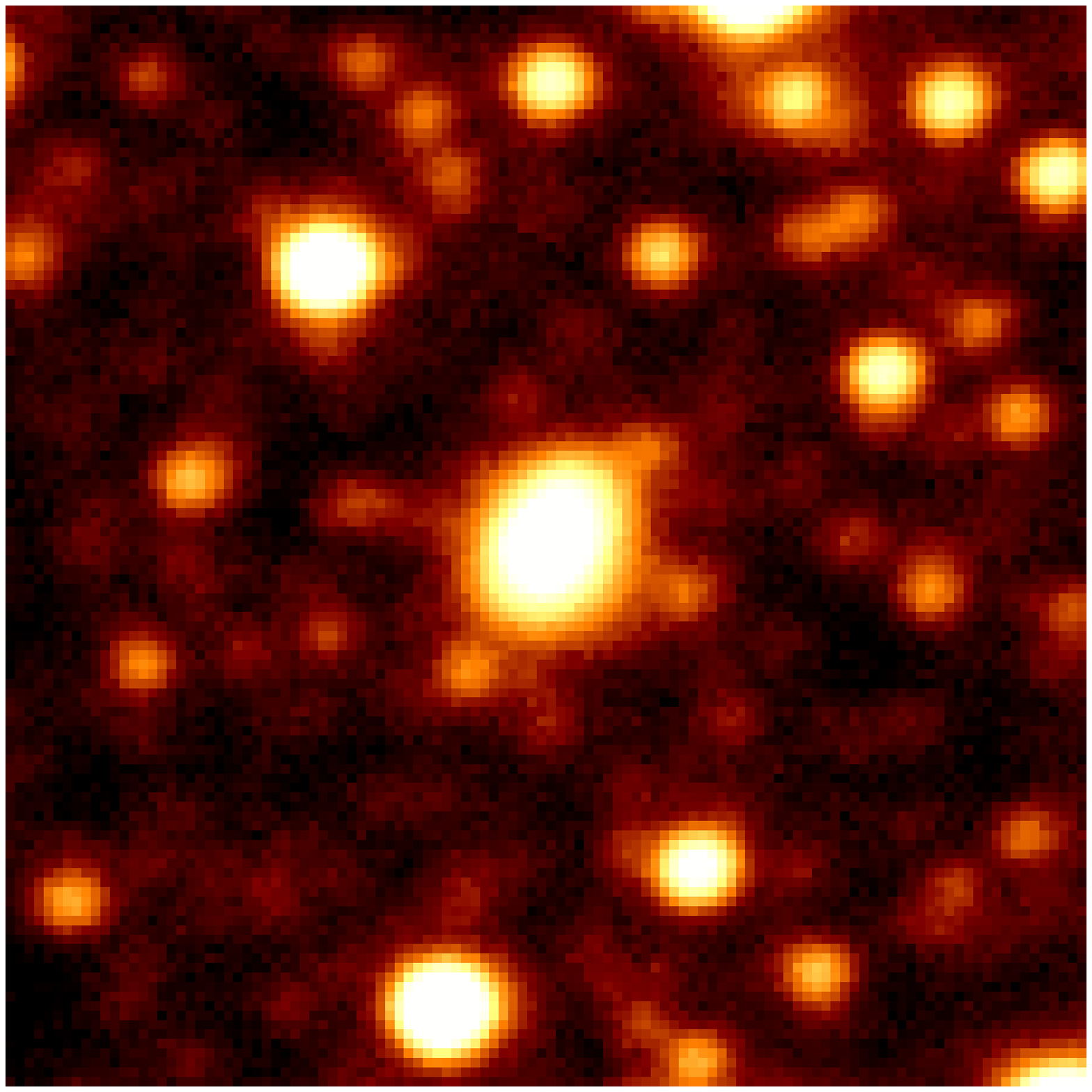}
   \end{center}
   \caption{OGLE-III $I$-band (left) and $V$-band (right) images of M~2-29. The same logarithmic scaling is applied in each 30$\times$30 arcsec$^2$ image where North is up and East is to the left. The contoured $I$-band inset highlights the elongated nebula oriented at a position angle (PA) of 168$^\circ$ that was missed by \hajduk.}
   \label{fig:images}
\end{figure}

\hajduks also claimed to find a short period of $\sim$23 days during the ingress seen in Fig. \ref{fig:lc}. Unfortunately this part of the M~2-29 lightcurve is very undersampled and \hajduks presented no quantitative evidence for periodicity. We detrended the same data and failed to find any significant periodicity in a Lomb periodogram (Fig. \ref{fig:fourier}). The strongest but statistically insignificant peak at 31.2 days in Fig. \ref{fig:fourier} is highly likely to be a sampling alias of order one month. A large component of the scatter appears to originate from the pipeline as there are large changes between the original and reprocessed lightcurves (inset Fig. \ref{fig:lc}). This further reduces the reliability of the data in this region. Low-level nebular contamination also needs to be considered for any contribution as artificial variability can be imprinted on non-stellar sources in the stellar-optimised OGLE reduction pipeline (Wo\'zniak 2000; Miszalski et al. 2009). Miszalski et al. (2009) showed lightcurves of some particularly bright PNe which showed artificial periods of one year. In M~2-29 the $I$-band nebular contamination is weak enough (Fig. \ref{fig:images}) to cause no problems for the pipeline but it is certainly not absent (\hajduk). 

\begin{figure}
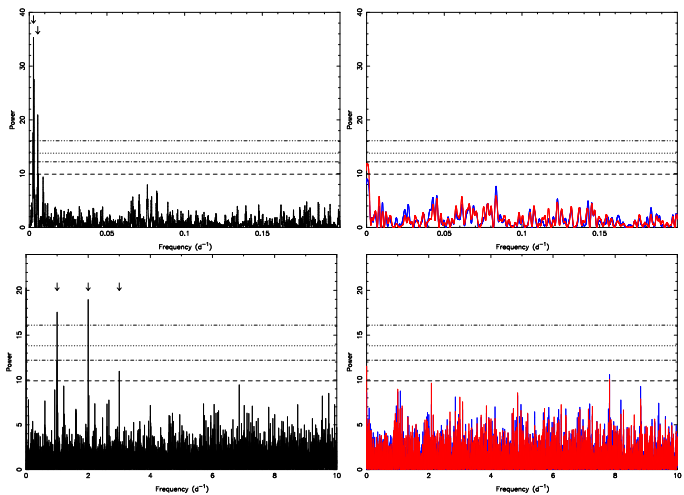

   \begin{center}
      \includegraphics[angle=270,scale=0.180]{fourierCBM2-29short.ps}
      \includegraphics[angle=270,scale=0.180]{fourierCBMACHOshort.ps}\\
      \includegraphics[angle=270,scale=0.180]{fourierCBM2-29.ps}
      \includegraphics[angle=270,scale=0.180]{fourierCBMACHO.ps}
   \end{center}
   \caption{Same as Fig. \ref{fig:fourier} but for the detrended OGLE-II and OGLE-III data at light maximum (left column) and the MACHO $V$ (blue) and $R$ (red) data during light minimum (right column). All marked peaks in the OGLE data are sampling aliases of 6 or 12 months, and 1, 2 or 3 days. These peaks are absent from the less frequently sampled MACHO data.}
   \label{fig:fourierCB}
\end{figure}

The purported appearance of periodic variability \emph{only} during ingress by \hajduks is also inconsistent with the litghtcurve. If this were the case we would expect variations to arise only from pauses in dust formation as in some WC9 stars (Veen et al. 1998). Otherwise, variability due to pulsations or a close companion would typically be expected to also occur in the optically thin portion of the lightcurve at light maximum. If a V651~Mon-like binary were present in M~2-29 we should be able to see this variability. Two well-sampled regions at light maximum containing 199 and 456 observations were examined for periodic variability (Fig. \ref{fig:lc}). Second order polynomials were fitted to each series and subtracted before rejoining the data for period analysis. Figure \ref{fig:fourierCB} shows no significant periodic variability present at light maximum. We also rule out the presence of pulsations above $\sigma=0.02$ mag with timescales of a few days or less which are somewhat common amongst Of(H) nuclei (Sec. \ref{sec:class}; Handler 1998, 2003).  
Although the MACHO data are of lesser quality compared to OGLE, we nonetheless searched for periods up to 1000 days at light minimum but found no significant variability (Fig. \ref{fig:fourierCB}).

\emph{We conclude that M~2-29 does not have a close irradiated binary companion with $P\la1$ day (e.g. Miszalski et al. 2009), nor does it have a close V651 Mon-like companion with $P\sim2$ weeks (e.g. M\'endez et al. 1982).} 

\subsection{OGLE $V$-band data}
\label{sec:vband}
Having ruled out a close binary companion, there is still a possibility that a wider companion may be present. Sparser OGLE $V$-band data for M~2-29 are well-suited to look for colour changes due to a companion (Zebrun 1998; Pigulski et al. 1993; Fig. 2 of \hajduk). The severe nebula contamination (Fig. \ref{fig:images}) leads us to suspect the pipeline processed data may be unreliable (Sects. 3.3 and 3.4 of Miszalski et al. 2009). Using the spectroscopy of P91 and P92 we find $V=15.5\pm0.1$ mag by comparing the flux at $\lambda$5480 to the equivalent in Vega. Remarkably, this magnitude agrees well with the $V$-band magnitudes at light maximum as remeasured by \hajduks despite the higher uncertainty with application of the difference image anaylsis method (Alard \& Lupton 1998) to the heavily nebula contaminated $V$-band data (Fig. \ref{fig:images}). This may be a result of the careful nebula subtraction by \hajduks and so we can trust their reprocessed $V$-band lightcurve. We must however assume that no deep minimum occured during the P91 and P92 epochs and this is reasonable given the large percentage of time spent at light maximum. 

The depths of the fading event are $I\sim$1.3 mag and $V\sim2.1$ mag. A colour change also occurs in $V-I$ from $0.7\pm0.1$ mag before minimum to $1.5$ mag during minimum. Could this be due to a companion? Assuming this change is solely caused by the dust produced during the event gives $A(V)=2.1$ mag. This corresponds to $A(I)\sim1.3$ assuming  $\lambda_\mathrm{eff}\sim8100$ \AA\ inferred from the mean extinction curve of Cardelli et al. (1989). This value of $A(I)$ is more suitable than $\lambda_\mathrm{eff}\sim9000$ \AA\ for the OGLE $I$-band since for $V-I<2$ mag there is negligible difference between the OGLE and Landolt systems (Udalski et al. 2002). As $A(I)$ matches the $I$-band depth of the fading event we cannot use the OGLE colours to infer the presence of a companion. Additional colour information or ideally NIR spectroscopy during the event would be required to investigate this further as the MACHO photometry are of insufficient quality.

\section{VLT FLAMES integral field unit spectroscopy of M~2-29}
\label{sec:obs}
\begin{figure}
   \begin{center}
      \includegraphics[scale=0.45,angle=270]{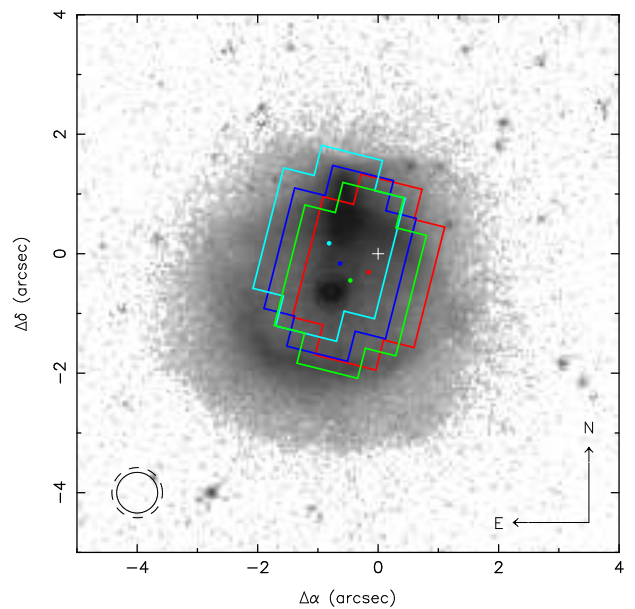}
   \end{center}
   \caption{
   The \emph{HST} F656N (H$\alpha$+[N~II]) image of M~2-29 (\hajduk) overlaid with approximate positions (dots) and footprints (lines) of the FLAMES mini-IFU for exposures in the grating settings LR01 (blue), LR02 (cyan), LR03 (green), and LR05, LR06 and LR07 (red). Differential atmospheric refraction shifts the $2 \times 3$\arcsec\ mini-IFU between each exposure and from the reported centre (white cross). The approximate mean seeing (FWHM) is shown by circles of diameters 0.68\arcsec\ (solid) and 0.84\arcsec\ (dashed) for non-LR03 and LR03 exposures, respectively.
   }
   \label{fig:overlay}
\end{figure}

Spectroscopic observations of M~2-29 were made by BM and AA on 11 June 2007 under visitor mode program 079.D-0764(A) with the FLAMES facility of the 8.2-m VLT UT2/Kueyen (Pasquini et al. 2002). 
Within the 25\arcmin\  diameter field of view FLAMES can deploy 15 miniature integral field units (mini-IFUs) to observe science targets, with another 15 reserved for sky measurements.
Each mini-IFU consists of an array of 20 square microlenses, each 0.52\arcsec\  a side, giving an effective rectangular aperture of $\sim$$2\times3$\arcsec\  with each corner being inactive. 
M~2-29 was observed with a mini-IFU centred at approximately $\alpha=18^\mathrm{h}06^\mathrm{m}40.85^\mathrm{s}$, $\delta=-26^\circ54'55.8''$ that was placed 9.9\arcmin\  from the centre of the FLAMES pointing at $\alpha=18^\mathrm{h}07^\mathrm{m}17.81^\mathrm{s}$, $\delta=-27^\circ00'25.3''$. 
Coordinates for the centre of M~2-29 were chosen from the SuperCOSMOS H$\alpha$ Survey (SHS; Parker et al. 2005) images which have a world coordinate system accurate to 0.3\arcsec\  rms. 
The non-configurable position angle (PA) of the mini-IFU was 165.84 degrees. At this orientation the mini-IFU favourably includes the central star and a substantial amount of the surrounding nebula emission (Fig. \ref{fig:overlay}). 

Table \ref{tab:log} summarises the exposures taken at light maximum (Fig. \ref{fig:lc}) with the GIRAFFE spectrograph equipped with the 2k$\times$4k EEV `Bruce' CCD. We used six of the eight available low-resolution grating settings: LR01, LR02, LR03, LR05, LR06 and LR07 to yield two continuous wavelength ranges of 3697--5074 \AA\  and 5722--8314 \AA. The wavelength range, resolving power $R$ ($\lambda/\Delta\lambda$) and exposure times made in each setting are given in Tab. \ref{tab:log}. Longer exposure times in bluer settings targeted the faint temperature sensitive diagnostic lines [O~III] $\lambda$4363 and [N~II] $\lambda$5755 required to measure high-quality chemical abundances (Sect. \ref{sec:chem}). Shorter exposures were not considered as no lines were found to be saturated in the longer exposures. The simultaneous arc lamp was turned off except for spectrophotometric standard star exposures. The average seeing conditions were 0.68$\pm$0.06\arcsec\  (FWHM) excluding the 1800 s LR03 exposure (0.84$\pm$0.19\arcsec). The airmass ranged from 1.83 to 1.03 with changes during exposures limited to $\sim$0.1 or less, with the exception of 0.39 for the 3600 s LR02 exposure. Sky transparency was generally poor with thin to thick clouds present (relative flux rms of 0.04--0.08).

\begin{table}
   \centering
   \caption{Log of M~2-29 VLT FLAMES GIRAFFE observations.}
   \begin{tabular}{lrrrr}
      \hline\hline
Grating setting & $\lambda$ (\AA) & $R$ ($\lambda/\Delta\lambda$)& Exp. (s)& UTC  \\
\hline                                                      
LR02 & 3948--4567 & 10200 &  600 & 01:17:40 \\
LR02 & 3948--4567 & 10200 & 3600 & 01:28:42 \\
LR05 & 5722--6510 & 11800 & 1800 & 02:29:44 \\
LR06 & 6420--7168 & 13700 &  900 & 03:01:03 \\  
LR06 & 6420--7168 & 13700 &  120 & 03:17:00 \\
LR07 & 7075--8314 &  8900 &  600 & 03:19:57 \\
LR01 & 3607--4085 & 12800 & 1800 & 03:31:08 \\
LR03 & 4486--5074 & 12000 & 1800 & 04:02:19 \\
LR03 & 4486--5074 & 12000 &  120 & 04:33:17 \\
\hline
   \end{tabular}
   \label{tab:log}
\end{table}

The science frames were reduced with the \textsc{IRAF}\footnote{IRAF is distributed by the National Optical Astronomy Observatories, which are operated by the Association of Universities for Research in Astronomy, Inc., under cooperative agreement with the National Science Foundation. See http://iraf.noao.edu} task \textsc{dofibers} after standard initial processing using \textsc{ccdproc}. We effectively removed the electronic glow defect of the `Bruce' CCD by including dark frames of the same duration in \textsc{ccdproc}. 
All science frames were cleaned of cosmic ray events (CREs) with the excellent L.A. Cosmic program (Van Dokkum et al. 2001). 
Aperture identifications were made in \textsc{dofibers} using fibre information provided in the OzPoz and FLAMES binary table extensions with particular care taken to not misidentify or misplace any apertures.
Low order polynomial fits were made to an average fibre flat field that was later divided through all individual flat field spectra. All science spectra were then divided by their corresponding normalised flat field spectrum. ThAr arc lamp spectra were also extracted and a wavelength solution was generated and applied to science spectra in the standard way using arc lists sourced from the ESO FLAMES pipeline. 

Flux calibration was provided by reducing in a similar fashion the sub-dwarf O star Feige 66 (Oke 1990) taken with the same field plate earlier the same night (albeit with a different IFU). All 20 fibres were averaged to form a combined Feige 66 spectrum from which the sensitivity function was derived. Spectra belonging to each grating setting were then flux calibrated individually. Determination of a smooth spectrophotometric response across all individual settings is discussed further in Sect. \ref{sec:chem}.

\begin{figure*}
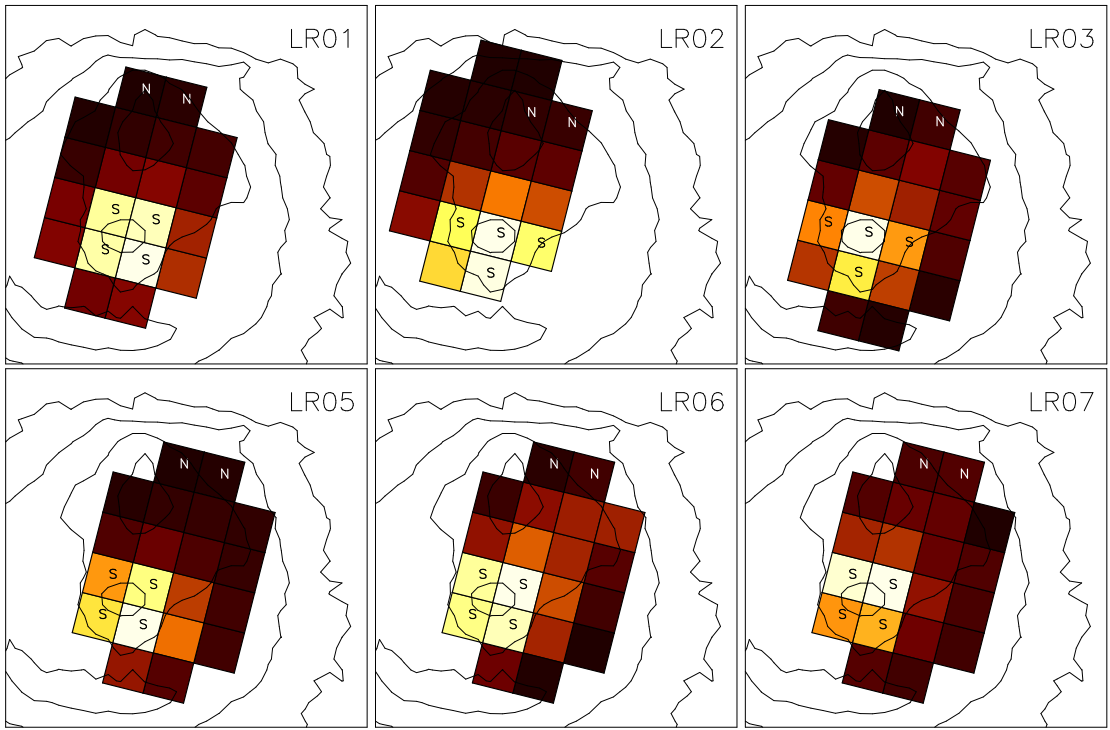

   \begin{center}
      \includegraphics[angle=270,scale=0.30]{ifu_LR01.ps}
      \includegraphics[angle=270,scale=0.30]{ifu_LR02.ps}
      \includegraphics[angle=270,scale=0.30]{ifu_LR03.ps}\\
      \includegraphics[angle=270,scale=0.30]{ifu_LR05.ps}
      \includegraphics[angle=270,scale=0.30]{ifu_LR06.ps}
      \includegraphics[angle=270,scale=0.30]{ifu_LR07.ps}
   \end{center}
   \caption{Selected spaxels averaged to form the CSPN (S) and outer nebula (N) spectra in each grating setting. Lighter colours indicate higher average intensity per spaxel reflecting the central star position. Contours are chosen to show the main nebula features (Fig. \ref{fig:overlay}).}
   \label{fig:selection}
\end{figure*}

Figure \ref{fig:selection} shows the spaxels selected for extraction of an average CSPN (S) and outer nebula (N) spectrum. In each grating setting an average sky spectrum is subtracted which comes from 15 dedicated sky fibres uniformly distributed in the 12.5\arcmin\  diameter FLAMES field-of-view. The selected CSPN spaxels were those four with the highest S/N in the continuum. There are multiple components to the CSPN spectrum including photospheric absorption and emission lines, a dense circumstellar nebula and the surrounding larger nebula. Obtaining a representative nebula spectrum for chemical abundance measurements unaffected by the dense stellar core is more difficult because of the small 2$\times$3\arcsec\  mini-IFU footprint. An ideal position would be the SE corner of the mini-IFU, however differential atmospheric refraction precludes coverage of blue wavelengths in this region. As a compromise we have selected an average of two spaxels per grating setting to cover the region on or just beyond the northern rim. 

\section{The central star of M~2-29}
\label{sec:cspn}
\subsection{Classification}
\label{sec:class}
\begin{figure*}
   \begin{center}
      \includegraphics[scale=0.5,angle=270]{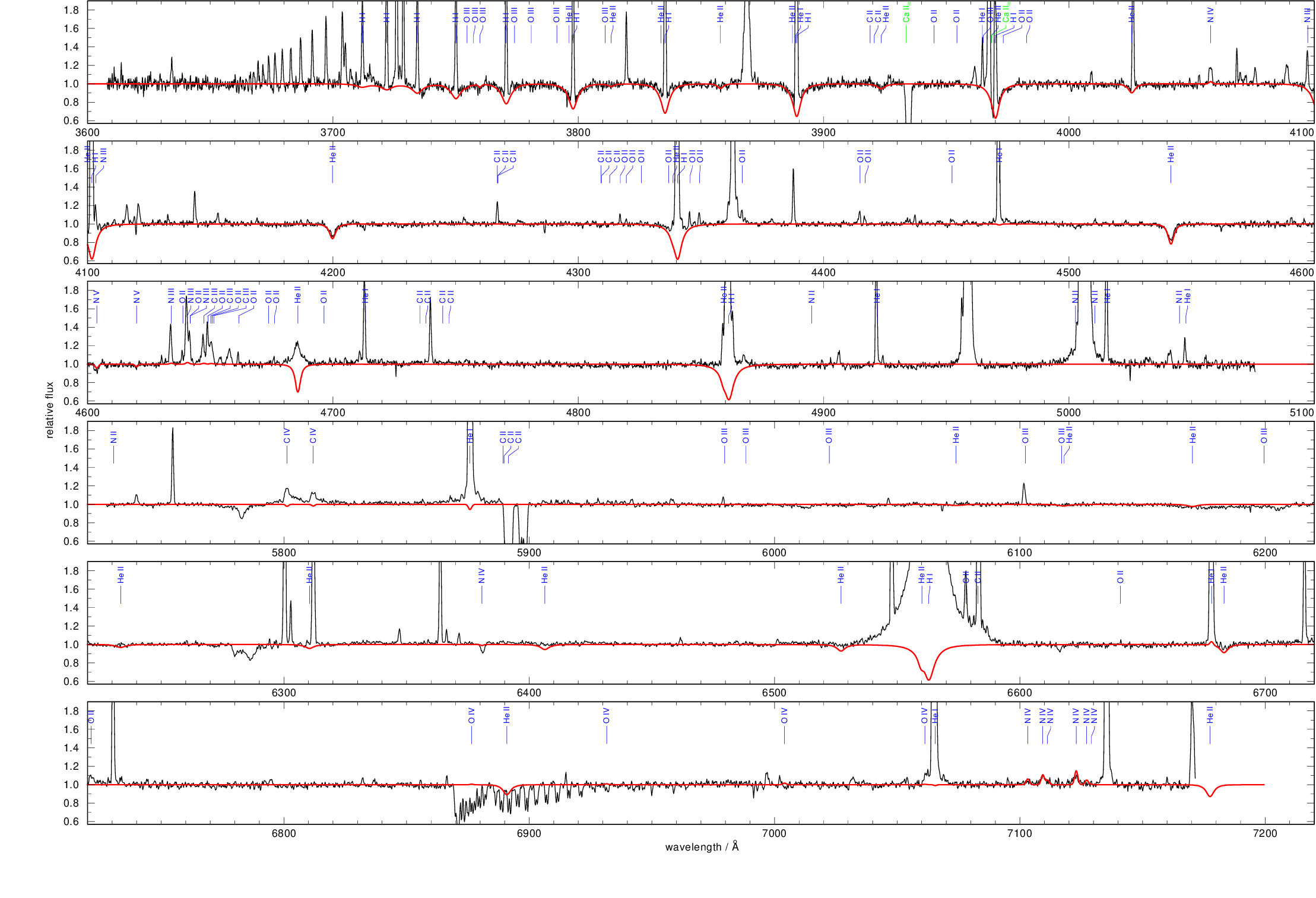}
   \end{center}
   \caption{TMAP NLTE model fit (red) of the rectified VLT FLAMES CSPN spectrum (black) with the main features identified.}
   \label{fig:thomas}
\end{figure*}

Figure \ref{fig:thomas} depicts our rectified spectrum of the CSPN with many common features identified. Overall the CSPN is a hot post-AGB star with He~II absorption lines as first shown by P91. Our derived and estimated properties of the CSPN are summarised in Tab. \ref{tab:cspn}. No features that could be assigned to a companion star are seen. A TMAP\footnote{http://astro.uni-tuebingen.de/$\sim$rauch/TMAP/TMAP.html} NLTE model atmosphere analysis (Werner et al. 2003; Rauch \& Deetjen 2003) was applied to yield $T_\mathrm{eff}=50\pm10$ kK from weak ionisation equilibrium of N~IV and N V lines and a surface gravity of log $g$=4.0$\pm$0.3 cm s$^{-2}$ from He~II line wings. Our $T_\mathrm{eff}$ is consistent with the total absence of nebular He~II emission and $T_\mathrm{eff}<40$ kK is unlikely as there are no He I absorption lines. The model normalised mass fractions used were close to the solar values: H (7.413$\times$10$^{-1}$), He (2.505$\times$10$^{-1}$), C (2.169$\times$10$^{-3}$), N (6.208$\times$10$^{-4}$) and O (5.379$\times$10$^{-3}$). An additional H-deficient TMAP model was also attempted but no adequate fit could be reached. With log $g$ known the distance to M~2-29 may be estimated following the procedure of Heber et al. (1984). We estimate the CSPN mass to be $0.60^{+0.28}_{-0.1}$ $M_\odot$ from the evolutionary tracks of Miller Bertolami \& Althaus et al. (2006) and take the monochromatic flux at $\lambda$5454 from our TMAP model. A distance of $6.9^{+2.7}_{-3.0}$ kpc is found where the errors are dominated by the uncertainty in log $g$.

\begin{table}
   \centering
   \caption{M~2-29 CSPN properties.}
   \label{tab:cspn}
   \begin{tabular}{lr}
      \hline\hline
      Property & Value \\
      \hline
      Spectral type & Of(H)\\
      $V$ & 15.50\\
      $T_\mathrm{eff}$ (kK) & $50\pm10$\\
      log $g$ (cm s$^{-2}$)& $4.0\pm0.3$ \\
      $M$ ($M_\odot$) & $0.60^{+0.28}_{-0.1}$\\
      Gravity distance (kpc) & 6.9$^{+2.7}_{-3.0}$\\
      $v_\infty$ (km s$^{-1}$) & $\la1200$ \\
      $\dot{M}$ ($M_\odot$ yr$^{-1}$) & $\sim$$10^{-6}$ \\
      \hline
   \end{tabular}
\end{table}

\begin{figure*}
   \begin{center}
      \includegraphics[angle=270,scale=0.35]{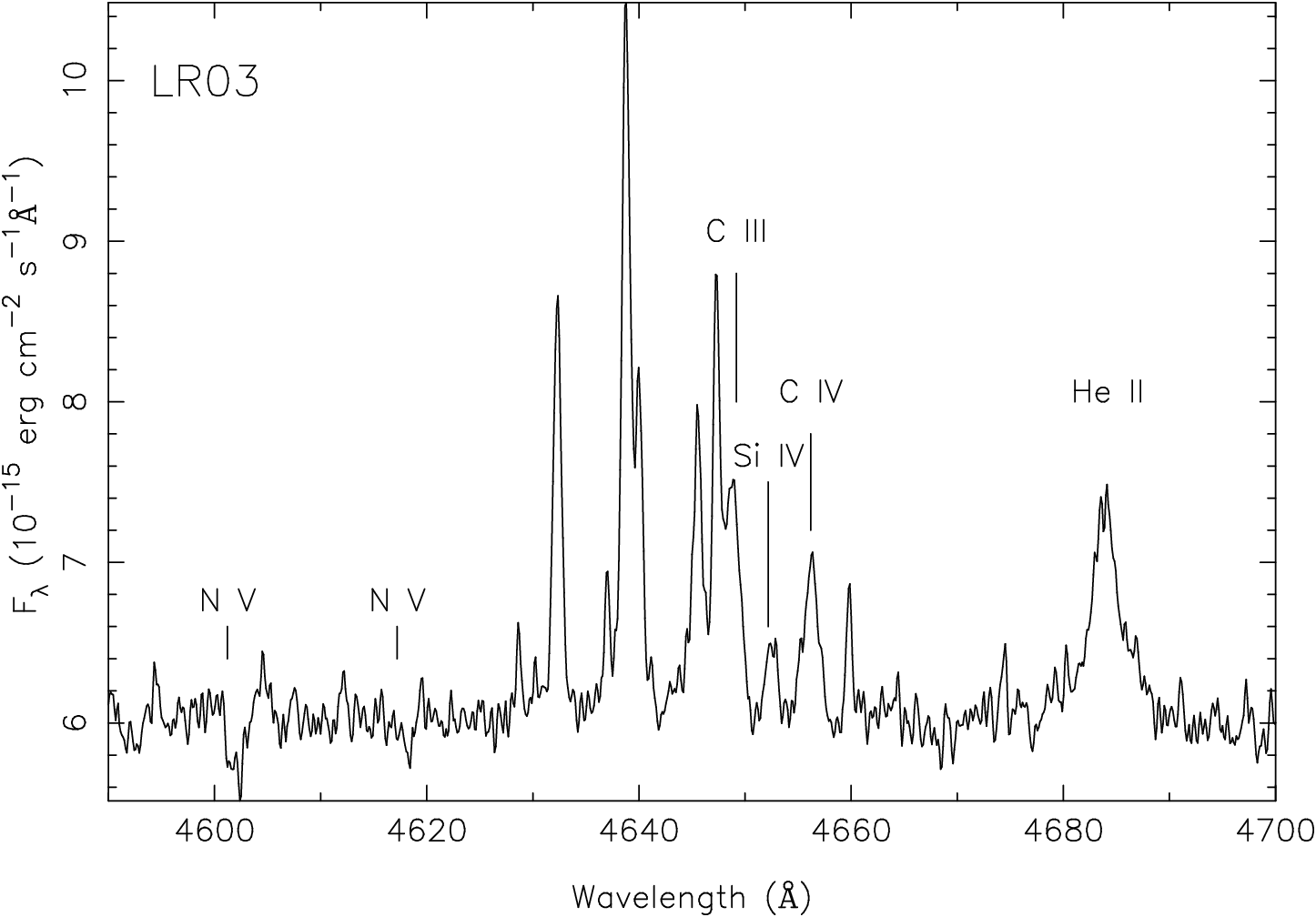}
      \includegraphics[angle=270,scale=0.35]{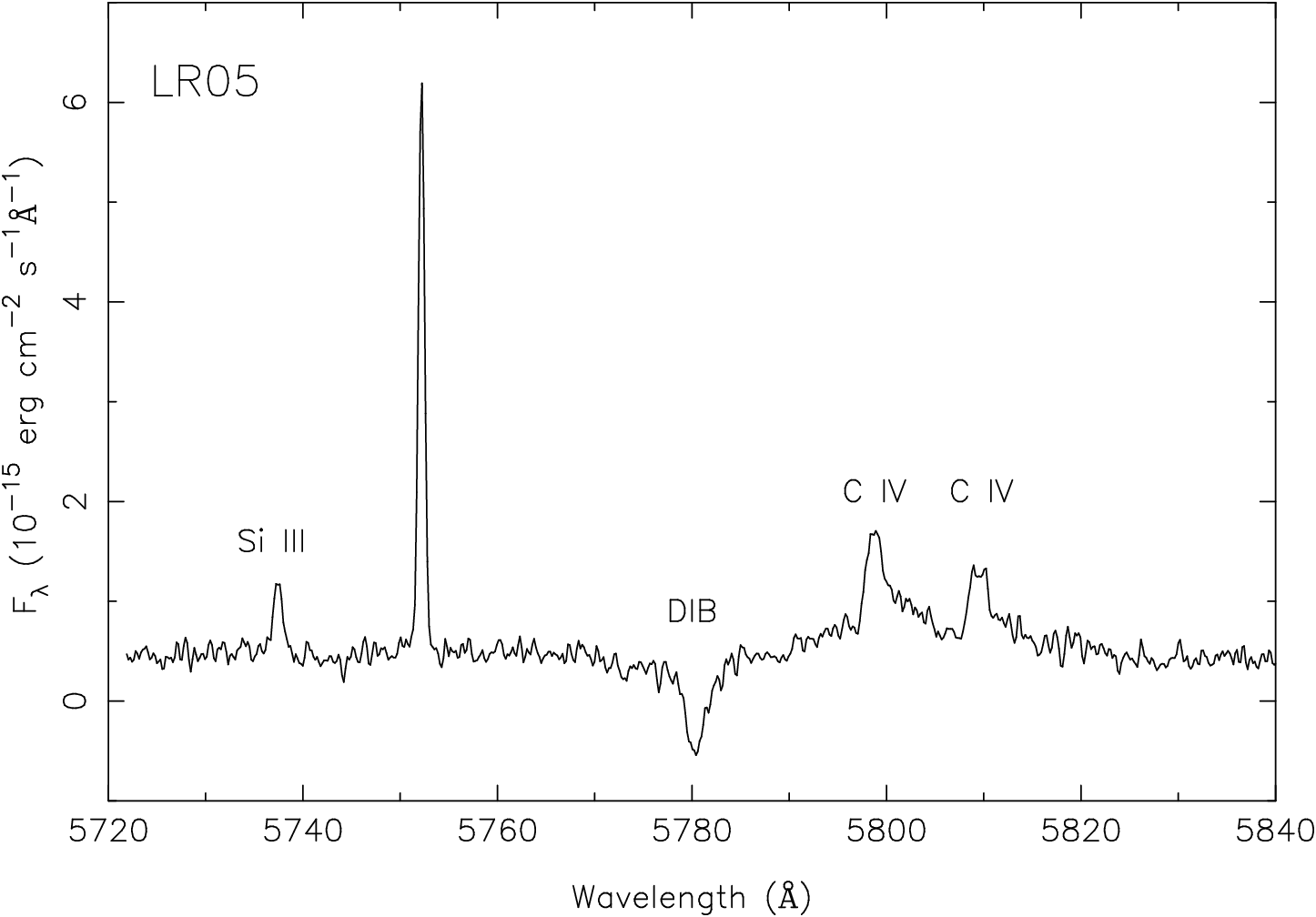}
   \end{center}
   \caption{FLAMES spectra of key regions with stellar emission lines and a diffuse interstellar band (DIB) labelled. Emission lines from the inner nebula of N~III, C~III, O~II (see Fig. \ref{fig:thomas} and M\'endez et al. 1990 for identifications) and [N~II] $\lambda$5755 are also present. 
   }
   \label{fig:emission}
\end{figure*}

Apart from the improved resolution of previously known stellar He~II and C~IV features (P91; P92), our deeper FLAMES spectra show many stellar emission lines of Si~II, Si~III, Si~IV, N~IV and N~V that are seen for the first time (Fig. \ref{fig:emission}). These features were identified as stellar because of their larger equivalent widths compared to the sharp, narrow nebular lines following Fig. 4 of M\'endez et al. (1990). 
Winds are not incorporated into the model so not all emission lines are fit. The peculiarly narrow C IV emission lines for example are relatively common in hot, windy post-AGB stars including $\sim$30\% of the symbiotic stars studied by Miko{\l}ajewska, Acker \& Stenholm (1997). The H-rich composition means that we can definitively rule out a [WR] classification. The narrow He~II $\lambda$4686 in emission with FWHM=$3.7\pm0.2$\ \AA\ and H-rich atmosphere classifies M~2-29 as Of(H) under the well-defined scheme of M\'endez et al. (1990) and M\'endez (1991). M~2-29 meets all the requirements for an Of classification, i.e. He~II $\lambda$4686 FWHM $\le4$\AA, non-blueshifted He~II $\lambda$4200, $\lambda$4540 in absorption and symmetric H$\gamma$ absorption. An upper limit to $v_\infty=1200$ km s$^{-1}$ for M~2-29 may be inferred from the spectroscopically similar NGC 6826 ($T_\mathrm{eff}=50\pm10$ kK, log $g$=3.9$\pm$0.2) which straddles the Of and higher $v_\infty$ Of-WR types (M\'endez et al. 1990; Pauldrach et al. 2004). 

 \subsection{The broad H$\alpha$ profile}
\begin{figure}
   \begin{center}
      \includegraphics[angle=270,scale=0.35]{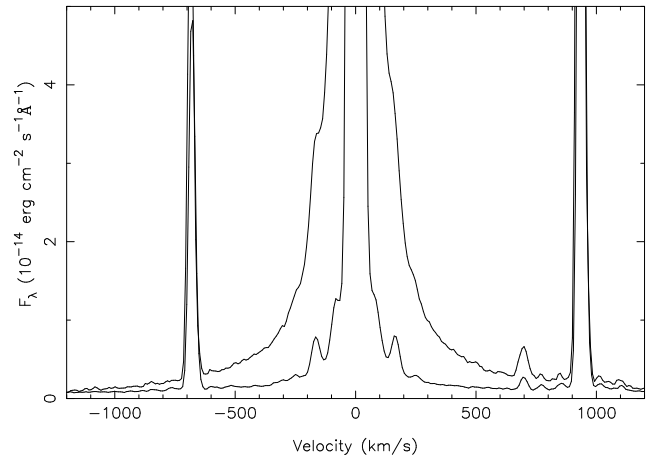}\\
   \end{center}
   \caption{H$\alpha$ profile for CSPN (top) and outer nebula (bottom). The top profile has been offset vertically to match the lower one.}
   \label{fig:halpha}
\end{figure}

Miszalski et al. (2009) first remarked upon the very broad H$\alpha$ profile which has a full-width at zero intensity (FWZI) of $\sim$2000 km s$^{-1}$ (Fig. \ref{fig:halpha}). Broad H$\alpha$ profiles are relatively uncommon in PNe with only a few known including the aforementioned NGC 6826 (e.g. Kudritzki et al. 1997, 2006; see also Kudritzki \& Puls 2000). Kudritzki et al. (1997, 2006) derived mass-loss rates of 10$^{-8}$--10$^{-7}$ $M_\odot$ yr$^{-1}$ for these CSPN by fitting their H$\alpha$ profiles. Larger mass-loss rates correspond to stronger H$\alpha$ profiles. In the case of M~2-29 we therefore estimate the mass-loss rate to be $\sim$10$^{-6}$ $M_\odot$ yr$^{-1}$ to produce the larger than normal H$\alpha$ profile for PNe. Our estimate requires confirmation through detailed modelling. It is encouraging that Gesicki et al. (2010) found the same rate is required to reproduce the kinematic age of the inner nebula, but they did not identify the CSPN as the responsible mass-losing star. As there appears to be no luminous AGB star companion in M~2-29 the mass-losing star must be the Of(H) CSPN.

Broad H$\alpha$ profiles are more commonly observed in symbiotic stars (Van Winckel et al. 1993; Munari \& Zwitter 2002) where they originate near the hot primary (Quiroga et al. 2002; Skopal 2006 and ref. therein). Skopal (2006) studied the H$\alpha$ profiles of symbiotic stars during their active and quiescent phases to find mass-loss rates of 10$^{-7}$--10$^{-6}$ $M_\odot$ yr$^{-1}$ and $\sim$$10^{-8}$ $M_\odot$ yr$^{-1}$, respectively. The geometry of the mass-loss was also modelled by Skopal (2006) and found to likely be in the form of an optically thin bipolar outflow. Gesicki et al. (2010) suggested such a geometry for M~2-29 consistent with this result. Section \ref{sec:sy} will discuss further similarities with symbiotic stars.

\subsection{Silicon emission lines} 
\label{sec:si}
There are three ionisation stages of Si visible in our deep FLAMES spectrum, e.g. Si II $\lambda$5041, $\lambda$5056, Si III $\lambda$4552, $\lambda$5739 (Fig. \ref{fig:emission}) and Si IV $\lambda$4654 (Fig. \ref{fig:emission}).
They are typically found in late [WC] type nuclei where they can reflect an extreme overabundance of Si (Leuenhagen \& Hamann 1998; Hultzsch et al. 2007). Hultzsch et al. (2007) modelled the spectrum of M 1-37 to find a Si overabundance of 28 times solar! There appears to be no explanation for this as Si is not expected to be enriched by AGB nucleosynthesis (Werner \& Herwig 2006). Leuenhagen \& Hamann (1998) suggested the overabundance may be evidence for additional mixing but could there be another explanation? 
The presence of a silicate-rich dust disk in M~2-29 (Sect. \ref{sec:dust}) could potentially explain this perceived overabundance. We speculate that the Si lines are formed via some interaction between the CSPN wind and the dust disk. Shielding supplied by the dust disk might help explain the presence of the low ionisation Si II species around the 50 kK CSPN. 

\subsection{Interstellar absorption features}
\label{sec:is}
Most absorption lines not fit by our TMAP model are diffuse interstellar bands (Herbig 1995). 
The Na I D1 and D2 absorption lines are resolved into three components each. Gesicki et al. (2010) claimed a component at $V_\mathrm{hel}=-100$ km s$^{-1}$ was caused by a superposition of the high velocity cloud (HVC) 002.9$-$06.2 that lies 3.4$^\circ$ away from M~2-29 with $V_\mathrm{lsr}=-88$ km s$^{-1}$ (Putman et al. 2002). The authors missed a closer HVC 004.8$-$06.2 (3.3$^\circ$ away) with $V_\mathrm{lsr}=+206$ km s$^{-1}$. As these clouds measure only $\sim$0.2$\times$0.2 degrees across on the sky there can be no line-of-sight superposition.

We measured equivalent widths (EW) of 0.82, 0.89 and 1.41 \AA\, for the Na I D1 lines, all of which lie in the saturated regime of the EW-$E(B-V)$ relation Munari \& Zwitter (1997). No estimate of $E(B-V)$ can therefore be made from the Na I D1 lines. Munari \& Zwitter (1997) offer K I $\lambda$7699 as an alternative indicator at higher values of $E(B-V)$, however our low resolving power and shallow exposure time in LR07 prevent its use here. Distance estimates from the Na I lines are also impractical towards the Galactic Bulge given the complex nature of randomly distributed dust clouds (K\"oppen \& Vergely 1998).

\section{Nebula properties of M~2-29}
\label{sec:neb}
\subsection{Reddening and integrated fluxes}
\label{sec:red}
The logarithmic extinction at H$\beta$, $c$, varies somewhat in the small amount of literature on M~2-29. Ground-based longslit spectroscopy values include 0.82 (Acker et al. 1992), 0.97 (P91; P92) and 1.18 (\hajduk). A value of 1.05 was measured from difficult to calibrate fibre spectroscopy (Exter et al. 2004), while the multiple grating settings of our FLAMES spectra make them unsuitable for measuring $c$. An independent measure of $c$ is required to check these values given the strong stellar H$\alpha$ emission from the CSPN that may have contaminated previous measurements. One method involves the use of suitable emission line fluxes of the whole nebula. This method includes an [O~III] to H$\beta$ conversion which requires the [O~III] and H$\alpha$ emission to be reasonably co-located. This is confirmed after inspection of the \emph{HST} images and helped by the small magnitude and variation of [N~II] ($\lambda$6548+$\lambda$6583)/H$\alpha$ across the nebula (0.05--0.2).

Table \ref{tab:fluxes} lists integrated H$\alpha$, H$\beta$ and [O~III] $\lambda$5007 fluxes of M~2-29.
Fluxes measured by Acker et al. (1992, A92) are longslit values scaled to the whole nebula. 
Other fluxes were measured directly from an [O~III] imaging survey of Galactic Bulge PNe (Kovacevic et al. 2011, K11), the Southern H$\alpha$ Sky Survey Atlas (SHASSA, Gaustad et al. 2001), and \emph{HST} F502N and F656N images (\hajduk). The \emph{HST} fluxes as listed exclude the CSPN, while values in parentheses include the CSPN. The H$\beta$ fluxes were calculated using [O~III] $\lambda$5007/H$\beta=4.86$ (an average of our FLAMES spectra) in all cases except Acker et al. (1992). The procedure for obtaining H$\alpha$ fluxes from aperture photometry of SHASSA data is described by Frew, Parker \& Russeil (2006).

For the HST fluxes we obtained the science-ready pipeline-reduced images from the MAST archive\footnote{http://archive.stsci.edu/hst}. Aperture photometry was performed with radii of 60 pixels (2.73\arcsec) and 8 pixels (0.36\arcsec) chosen to cover the whole nebula (including CSPN and inner nebula) and the CSPN and inner nebula, respectively. Surrounding annuli of width 10 and 2 pixels were used to subtract the sky and nebula contributions in each case. Flux calculations followed the WFPC2 Photometry Cookbook\footnote{http://www.stsci.edu/hst/wfpc2/wfpc2\_faq/wfpc2\_nrw\_phot\_faq.html} that improves upon Holtzman et al. (1995). Known \emph{HST} sensitivities and rectangular filter widths of 35.82 \AA\ (F502N) and 28.33 \AA\ (F656N) were used to find the final sky-subtracted flux in each aperture. Both SHASSA and \emph{HST} H$\alpha$ fluxes were adjusted  by 0.01 dex to remove the small ($<2$\%) contribution from [N~II] emission in the filter bandpasses. The \emph{HST} fluxes appear to be systematically fainter than ground-based measurements which may be explained by the higher spatial resolution. 

Since the same aperture was used in each image we consider the H$\alpha$ to [O~III] ratio to be reliable. We therefore find $c=0.95\pm0.05$ from the average \emph{HST} derived value of H$\alpha$/H$\beta=5.82$ (Tab. \ref{tab:fluxes}). This value corresponds to $E(B-V)=0.65$ or $A_V=2.0$ and is free of any contribution from the CSPN or inner nebula. Table \ref{tab:fluxes} also explains the higher $c=1.18$ measured by HZG08. Comparing the CSPN-inclusive H$\alpha$ flux to the CSPN-free H$\beta$ flux gives H$\alpha$/H$\beta=6.92$ or $c=1.18$. The stellar H$\alpha$ emission is clearly distorting the value obtained by HZG08. The higher quality of the P91 spectra could explain the better agreement with their value of $c=0.97$.

\begin{table}
   \centering
   \caption{Integrated fluxes of M~2-29 uncorrected for reddening.}
   \label{tab:fluxes}
   \begin{tabular}{lcccc}
      \hline\hline
      Line & A92 &  K11 & SHASSA & \emph{HST} \\
      \hline
      {}$-$log F(H$\beta$) & 12.18 &  12.14 & - &  12.34 (12.27) \\  
      {}$-$log F([O~III]) & 11.47 &  11.45 &- & 11.65 (11.59)\\
      {}$-$log F(H$\alpha$) & 11.46 & - & 11.50  & 11.58 (11.50)\\
      \hline
   \end{tabular}
\end{table}
\subsection{Distance and ionised mass}
With our reddening of $c=0.95\pm0.05$, measured angular diameter of $5.0\times4.8$ arcsec at 10\% of peak intensity and an integrated H$\alpha$ flux of $-$log F(H$\alpha$) = 11.58 (Tab. \ref{tab:fluxes}) we estimated the distance to M~2-29 using the H$\alpha$ surface brightness-radius relation (SBR) of Frew \& Parker (2006). The average relation suitable for M~2-29 gives a distance of $7.7\pm2.3$ kpc. We adopt a final distance of $7.4\pm1.8$ kpc which is a weighted mean of the gravity distance (Sect. \ref{sec:class}) and the SBR distance. M~2-29 is therefore a Galactic Bulge PN entirely consistent with its Galactic coordinates (PN G004.0$-$03.0) and small radius.

The ionised mass of the nebula can now be calculated which can generally serve as a useful discriminant between PNe and symbiotic nebulae. Santander-Garc{\'i}a, Corradi \& Mampaso (2007) note typical ionised masses of 10$^{-4}$--10$^{-2}$ $M_\odot$ for symbiotic nebulae and 0.1--1.0 $M_\odot$ for PNe. Frew \& Parker (2010) give a range of 0.005--3 $M_\odot$ derived from a larger sample of PNe. Symbiotic nebulae typically have smaller masses since the nebula forms from the tenuous wind of the evolved companion (see Corradi 2003 for further discussion), but this method is not foolproof because some symbiotic nebulae can reach 0.1 $M_\odot$ (Santander-Garc{\'i}a et al. 2008). Using the formula of Pottasch (1996):
\begin{equation}
   \mathrm{M}_\mathrm{neb} (\mathrm{M}_\odot) = 4.03 \times 10^{-4} \epsilon^{1/2} d^{5/2} \mathrm{H}\beta^{1/2} \theta^{3/2}
   \label{eqn:mion}
\end{equation}
where $d$ is the distance ($7.4\pm1.8$ kpc), H$\beta$ is the dereddened flux ($0.46\times10^{-11}$ erg cm$^{-2}$ s$^{-1}$), $\theta$ is the nebular radius (4.9\arcsec), and $\epsilon=0.3$ is the volume filling factor. This gives M$_\mathrm{neb}=0.08\pm0.05$ M$_\odot$ which lies in between a symbiotic nebula and PN but is more consistent with a PN. The nebula does not resemble those around D-type symbiotic stars (e.g. Corradi 2003), however the multiple nebulae present are shared with D'-type symbiotics in which the outer nebula is presumably ejected by the white dwarf (see Sect. \ref{sec:sy}). There is therefore no reason to suspect the nebula around M~2-29 is not a \emph{planetary} nebula. 

\subsection{Chemical abundances}
\label{sec:chem}
M~2-29 has long been considered a halo PN because of its low oxygen abundance of $7.4\pm0.1$ dex derived from ground-based longslit spectroscopy (P91; Exter et al. 2004). This anomalously low abundance is \emph{entirely} the result of low spatial resolution ground-based observations averaging over the strong spatial variation of [O~III] $\lambda$4363/$\lambda$5007 (TP97; Miszalski et al. 2009). TP97 measured an O abundance of 8.5 dex from \emph{HST} spectroscopy of the inner nebula and yet it is still incorrectly referred to as a halo PN in the literature (Otsuka et al. 2009; Lutz et al. 2010; Stasi{\'n}ska et al. 2010; Sch\"onberner et al. 2010). To settle the issue we have recalculated the chemical abundances of M~2-29 from our spectrum of the outer nebula.

Figure \ref{fig:outer} displays the joined FLAMES spectra of the outer nebula and the inner nebula. Note the dramatic difference between the outer nebula and the core which shows much stronger [Ne~III], He~I and [O~III] 4363 emission lines in addition to lines not seen in the outer nebula (e.g. [K IV] $\lambda$6102, [Cl IV] $\lambda$7530, 8045). Here we focus only on the outer nebula for which it is more straight-forward to calculate abundances. Even after flux calibration of the six individual spectra there are still some undetermined offsets between them. These offsets were found by fixing $c=0.95$ (Sect. \ref{sec:red}) and fitting the H~I and He~I emission lines to their expected Case B intensities assuming a Howarth (1983) reddening law. Figure \ref{fig:caseB} shows the results for H~I and demonstrates the corrected spectrophotometric solution is suitable for abundance determination. Table \ref{tab:emlines} lists our corrected emission lines intensities before and after dereddening. 

\begin{figure*}
   \begin{center}
      \includegraphics[scale=0.80,angle=270]{outerNeb.ps}\\
      \includegraphics[scale=0.80,angle=270]{innerNeb.ps}
   \end{center}
   \caption{VLT FLAMES spectra of the outer (top) and inner (bottom) nebula. }
   \label{fig:outer}
\end{figure*}

\begin{figure}
   \begin{center}
      \includegraphics[angle=270,scale=0.35]{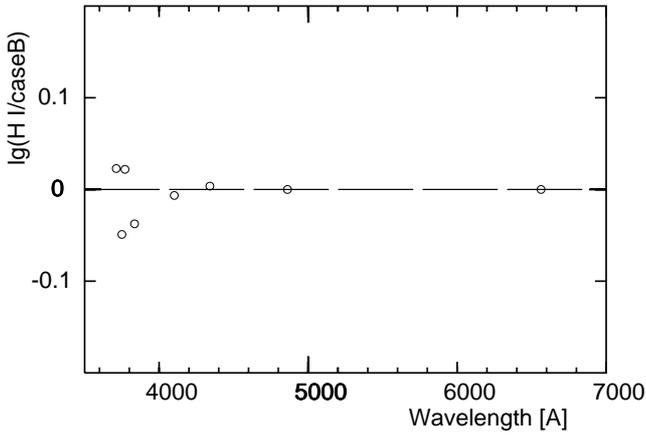}
   \end{center}
   \caption{Residual H~I emission line intensities for Case B conditions and $c=0.95$. The slightly larger scatter at $\lambda<4000$\AA\  is consistent with the lower intensities of these lines and the difficulty of instrumental and flux calibration at these wavelengths.}
   \label{fig:caseB}
\end{figure}

The measured line intensities were analysed via the plasma diagnostic program HOPPLA (Acker et al. 1991; K\"oppen et al. 1991; see also Girard et al. 2007). The extinction constant was fixed to $c=0.95$ and an iterative process was used to consistently derive the electron temperatures of $T_e$([N~II])=8980 K, $T_e$([O~III])=10500 K and an electron density $n_e=2220$ cm$^{-3}$. Application of the usual ionisation correction factors allowed for the derivation of the elemental abundances in the format $12+\log(X/H)$ in Tab. \ref{tab:abundances}. Table \ref{tab:abundances} includes for comparison abundance measurements of M~2-29 from P91, TP97, Exter et al. (2004, E04), as well as average abundances for Bulge PNe (Cuisinier et al. 2000) and Solar abundances (Asplund et al. 2005). Overall the abundances are in good agreement with TP97 except for neon. Our measured oxygen abundance of 8.3 dex is low but entirely consistent at the 2$\sigma$ level with an average Bulge PN (Cuisinier et al. 2000). There is now no reason to consider M~2-29 as a Halo PN since the oxygen abundance is only $0.4$ dex lower than Solar.

\begin{table}
   \centering
   \caption{Emission line intensities of the outer nebula}
   \label{tab:emlines}
   \begin{tabular}{lrr}
      \hline\hline
        Identification & $I(\lambda)$ & $I_0(\lambda)$ \\
        \hline
        {}H I 3712.0          &  0.93 & 1.64 \\                                     
        {}[S III] 3721.9      &  1.85 & 3.24 \\                                    
        {}[O II] 3726.0       &  21.28 & 37.32 \\                                   
        {}[O II] 3728.7       &  11.90 & 20.85 \\                                   
        {}H I 3734.4          &  1.55 & 2.71 \\                                    
        {}H I 3750.1          &  1.66 & 2.89 \\                                    
        {}H I 3770.6          &  2.05 & 3.53 \\                                    
        {}[S III] 3797.8      &  2.76 & 4.70 \\                                    
        {}He I 3819.6         &  0.80 & 1.35 \\                                    
        {}H I 3835.4          &  4.03 & 6.77 \\                                    
        {}[Ne III] 3868.7     &  14.62 & 24.20 \\                                   
        {}He I 3889.0         &  10.47 & 17.19 \\                                   
        {}He I 3964.7         &  0.88 & 1.39 \\                                    
        {}[Ne III] 3967.4     &  4.45 & 7.04 \\                                    
        {}H I 3970.1          &  1.71 & 2.71 \\                                    
        {}He I 4009.2         &  0.22 & 0.35 \\                                    
        {}He I 4026.2         &  1.30 & 2.00 \\                                    
        {}H$\delta$ 4101.7    &  17.13 & 25.49 \\                                         
        {}O II 4120.8         &  0.17 & 0.25 \\                                    
        {}He I 4143.8         &  0.21 & 0.31 \\                                    
        {}C II 4267.0         &  0.22 & 0.30 \\                                    
        {}H$\gamma$ 4340.5    &  35.99 & 47.48 \\                                         
        {}O II 4345.6         &  0.10 & 0.13 \\                                    
        {}O II 4349.4         &  0.11 & 0.14 \\                                    
        {}[O III] 4363.1      &  3.01 & 3.92 \\                                    
        {}O II 4366.9         &  0.08 & 0.11 \\                                    
        {}He I 4387.9         &  0.58 & 0.75 \\                                    
        {}He I 4471.5         &  4.59 & 5.66 \\                                    
        {}He I 4713.2         &  0.65 & 0.70 \\                                    
        {}H$\beta$ 4861.3     &  100.00 & 100.00 \\                                      
        {}He I 4921.9         &  1.51 & 1.46 \\                                    
        {}[O III] 4958.8      &  176.03 & 167.05 \\                                 
        {}[O III] 5006.7      &  562.33 & 520.10 \\                                 
        {}He I 5015.7         &  2.97 & 2.73 \\                                    
        {}[N II] 5754.5       &  0.71 & 0.46 \\                                    
        {}He I 5875.7         &  21.25 & 13.29 \\                                  
        {}[O I] 6300.1        &  1.75 & 0.95 \\                                    
        {}[S III] 6312.0      &  1.44 & 0.78 \\                                    
        {}[O I] 6363.6        &  0.62 & 0.33 \\                                    
        {}[N II] 6547.9       &  26.81 & 13.37 \\                                  
        {}H$\alpha$ 6562.8    &  578.63 & 287.28 \\                                       
        {}C II 6578.0         &  0.81 & 0.40 \\                                    
        {}[N II] 6583.3       &  84.53 & 41.70 \\                                  
        {}He I 6678.2         &  9.09 & 4.35 \\                                    
        {}[S II] 6716.4       &  4.48 & 2.12 \\                                    
        {}[S II] 6731.6       &  7.46 & 3.52 \\                                    
        {}He I  7065.3        &  10.53 & 4.51 \\                                   
        {}[Ar III] 7135.6     &  27.62 & 11.61 \\                                   
        {}He I 7281.4          & 1.77 & 0.72 \\                                      
        {}[O II] 7320.0       &  4.62 & 1.85 \\
        {}[O II] 7330.0       &  3.97 & 1.59 \\                                    
        {}[Ar III] 7750.9     &  3.95 & 1.43 \\                                    
        \hline                 
   \end{tabular}
\end{table}

\begin{table}
   \centering
   \caption{Chemical abundances of M~2-29.}
   \label{tab:abundances}
   \begin{tabular}{lrrrrrrrr}
      \hline\hline
      Element & P91 & TP97 & E04 & this work & Bulge &  Solar\\                    
      \hline
      He      & 10.97 & -  &11.00& \textbf{11.02} & 10.90    & 10.93\\
      N       & 7.2   & 7.9  &7.2& \textbf{7.8}  & 8.2      & 7.8\\
      O       & 7.3   & 8.5  &7.5& \textbf{8.3}  & 8.7      & 8.7\\
      Ne      & 6.7   & 8.0  &6.8& \textbf{7.4}  & -        & 7.8\\
      S       & 5.9   & 6.5  &5.6& \textbf{6.6}  & 6.9      & 7.1\\
      Ar      & 5.3   & -    &5.8& \textbf{6.0}  & 6.2          & 6.2\\
      \hline
      log(N/O) & $-$0.1 & $-$0.6& $-$0.3 & \textbf{$-$0.5} & $-$0.5 & $-$0.9\\ 
      log(Ne/O)& $-$0.6 & $-$0.5& $-$0.7 & \textbf{$-$0.9} & -      & $-$0.9\\
      log(S/O) & $-$1.4 & $-$2.0& $-$1.9 & \textbf{$-$1.7} & $-$1.8 & $-$1.6\\
      log(Ar/O)& $-$2.0 & -     & $-$1.7 & \textbf{$-$2.3} & $-$2.5 & $-$2.5\\
      \hline
   \end{tabular}
\end{table}

\section{Dust properties of M~2-29}
\label{sec:dust}
\begin{figure}
   \begin{center}
      \includegraphics[angle=270,scale=0.36]{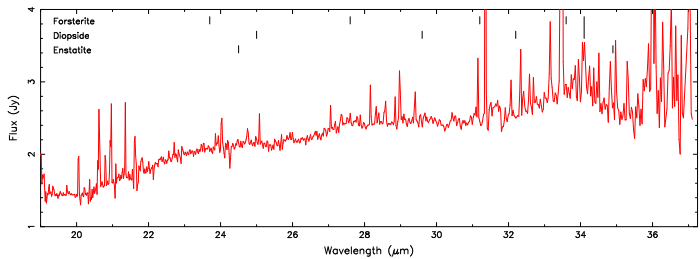}
   \end{center}
   \caption{Archival \emph{Spitzer} IRS spectrum of M~2-29. Crystalline silicate features are identified (see text) and the IRS spectrum at bluer wavelengths can be found in Gesicki et al. (2010).}
   \label{fig:odust}
\end{figure}

The large 70 AU radius at which dust forms around M~2-29 suggests the presence of a dust disk similar to \cpds (Sect. \ref{sec:lc}) and this is also supported by NIR and MIR observations. Gesicki et al. (2010) showed an archival \emph{Spitzer} IRS spectrum of M~2-29 covering 5.7--14.2 $\mu$m. A peak at 11.3 $\mu$m was identified by the authors as a polycyclic aromatic hydrocarbon (PAH) band, but the absence of other bands at 6.2, 7.7 and 8.7 $\mu$m is inconsistent with this interpretation. The same feature is detected in the D'-type symbiotic HD 330036 by Roche, Allen \& Aitken (1983) which they attributed to SiC dust. Figure \ref{fig:odust} shows the IRS spectrum at longer wavelengths after basic pipeline calibrations and reductions were applied.  
To identify crystalline silicate features we compared the IRS spectrum of M~2-29 with the features identified in the ISO spectrum of NGC 6302 (Tab. 1 of Molster et al. 2001) and in a set of evolved stars (Tab. 1 of Molster et al. 2002). We note the clear presence of the following minerals (features with S/N above 30): Forsterite (23.7, 27.6, 31.2 and 33.6 $\mu$m), Diopside (25.0, 29.6 and 32.2 $\mu$m), Enstatite (24.5 and 34.9 $\mu$m) and a feature at 34.1 $\mu$m described as a combination of Forsterite and Diopside. 
The silicate rich composition alone is not enough for a dual-dust chemistry (Cohen et al. 2002) which cannot be assigned to M~2-29 as there are no strong PAH bands at bluer MIR wavelengths. 

Table \ref{tab:mags} contains available near-, mid- and far-infrared fluxes from many ground and space-based missions (Beichman et al. 1988; Price et al. 2001; Skrutskie et al. 2006; Churchwell et al. 2009; Carey et al. 2009; Ishihara et al. 2010). The core of M~2-29 as detected by 2MASS has colours $J-H=0.41$ and $H-K_s=0.17$, well within the region occupied by PNe and substantially below the area that characterizes the D-type symbiotics (e.g. Corradi et al. 2008). 
The steepness of the optical-NIR spectral energy distribution excludes a
major contribution by free-free emission from the wind of the hot CSPN. We have therefore subtracted the stellar photospheric contribution
which we normalized in the optical. The Rayleigh-Jeans tail of this
starlight effectively removes any possibility of a credible excess of NIR
emission (i.e. from a stellar companion), even depending on 2MASS data alone. 
But we emphasise this does not rule out the existence of a companion of sub-giant luminosity or fainter.

\begin{table}
   \centering
   \caption{NIR, MIR and FIR fluxes of M~2-29 uncorrected for reddening.}
   \label{tab:mags}
   \begin{tabular}{lrr}
      \hline\hline
      Band & $\lambda_\mathrm{cen}$ ($\mu$m) & Flux (mJy) \\
      \hline
      2MASS $J$ & 1.23          & 4.4                 \\  
      2MASS $H$ & 1.66          & 4.1                \\   
      2MASS $K_s$ & 2.16        & 7.9                \\  
      IRAC [3.6] & 3.56         & 28.9                \\  
      IRAC [4.5] & 4.52         & 40.5                \\ 
      IRAC [5.8] & 5.73         & 36.3                \\  
      IRAC [8.0] & 7.91         & 71.2                \\ 
      \emph{MSX} $A$ & 8.28     & 151                      \\  
      \emph{MSX} $C$ & 12.13    & 227                   \\  
      \emph{MSX} $D$ & 14.65    & 216                   \\ 
      \emph{MSX} $E$ & 21.34    & 186                   \\
\emph{AKARI} S18 & 18.0                  & 1180                  \\
      MIPS [24] & 23.7                  & 1150                  \\ 
      \emph{IRAS} [12] & 12.0                  & 850:                 \\
      \emph{IRAS} [25] & 25.0                  & 2000                  \\
      \emph{IRAS} [60] & 60.0                  & 2280                   \\
      \emph{IRAS} [100] & 100.0                & $<$75000               \\
      \hline
   \end{tabular}
\end{table}

\begin{figure*}
   \begin{center}
      \includegraphics[scale=0.31]{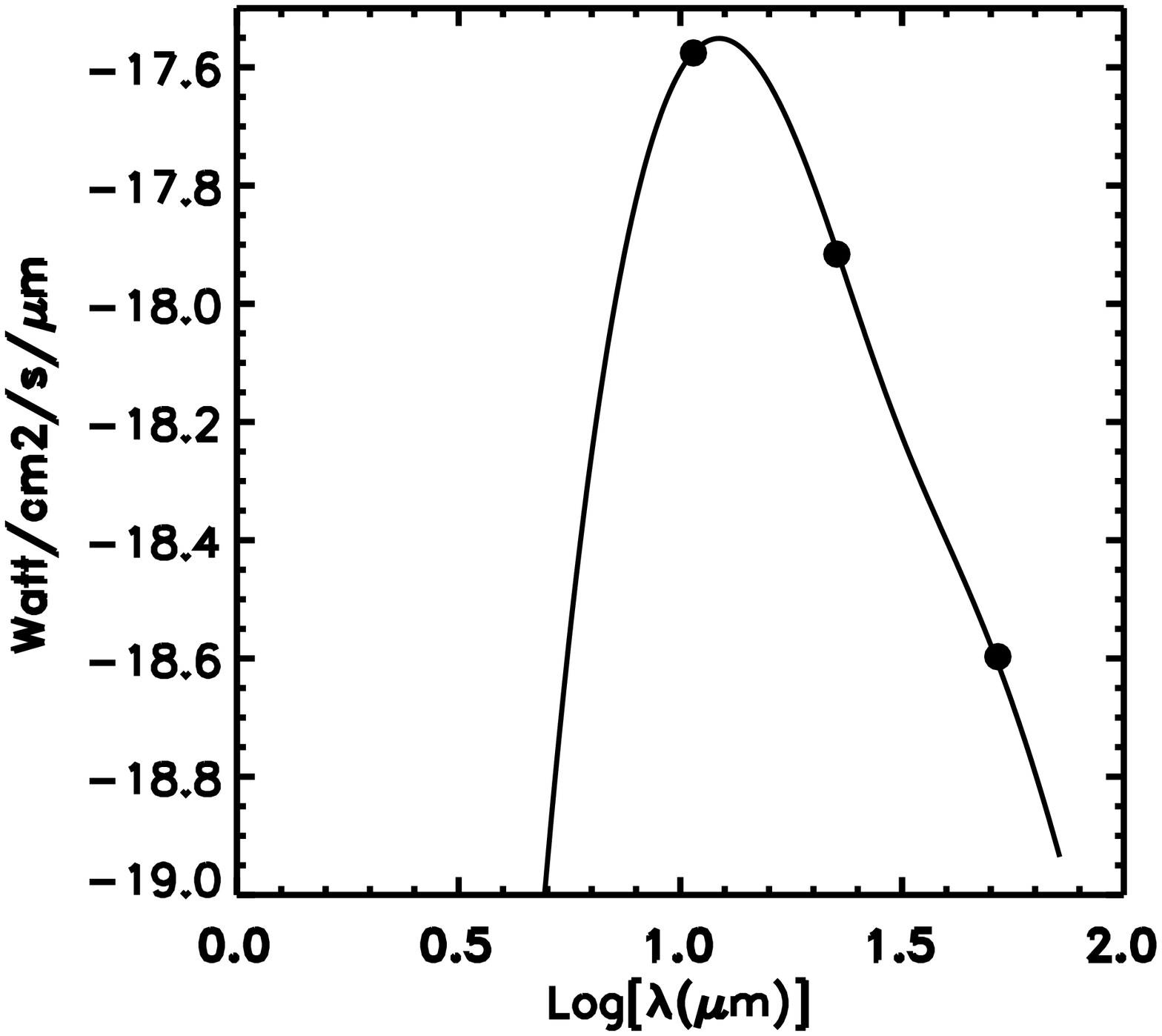}
      \includegraphics[scale=0.31]{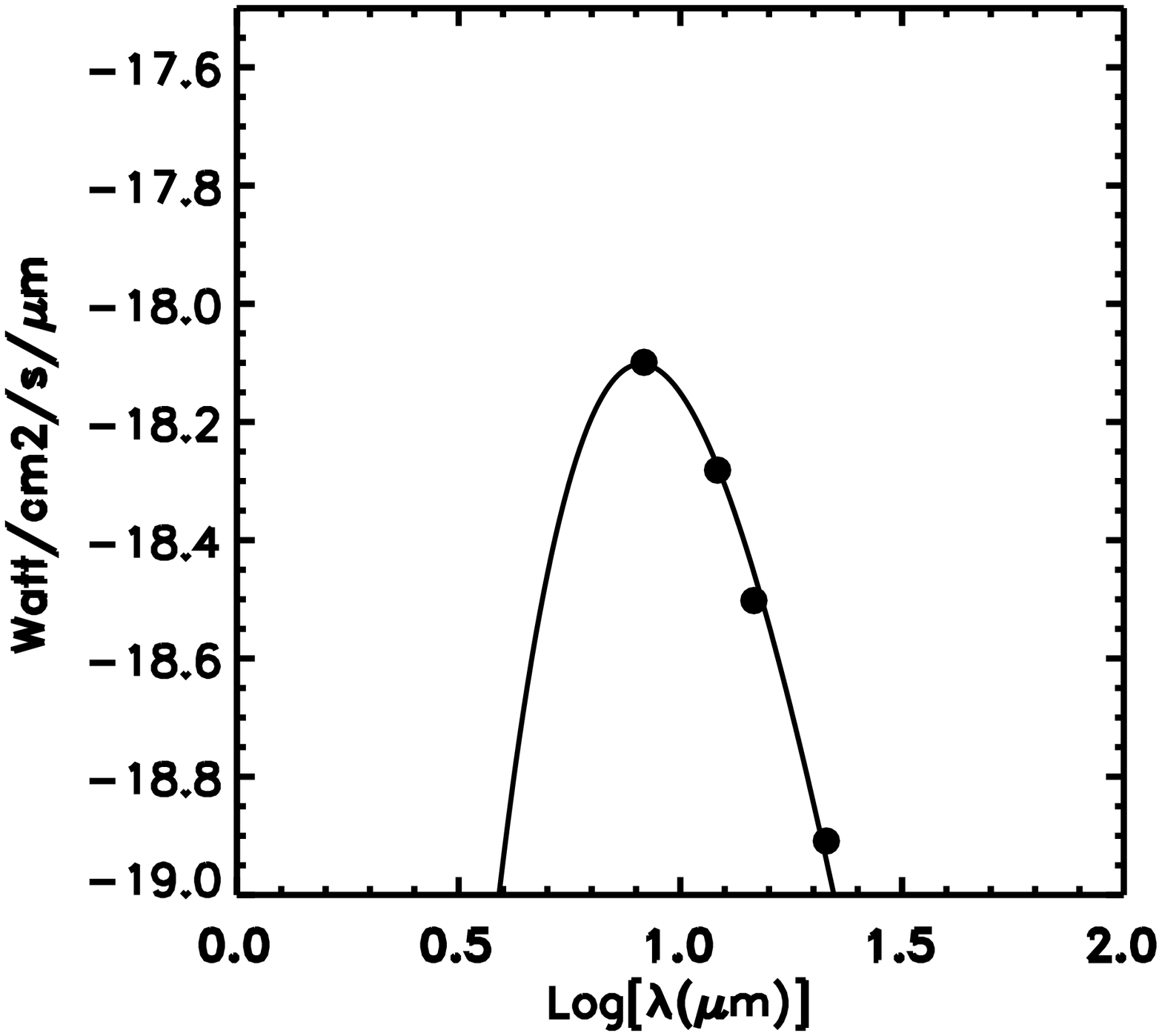}
      \includegraphics[scale=0.31]{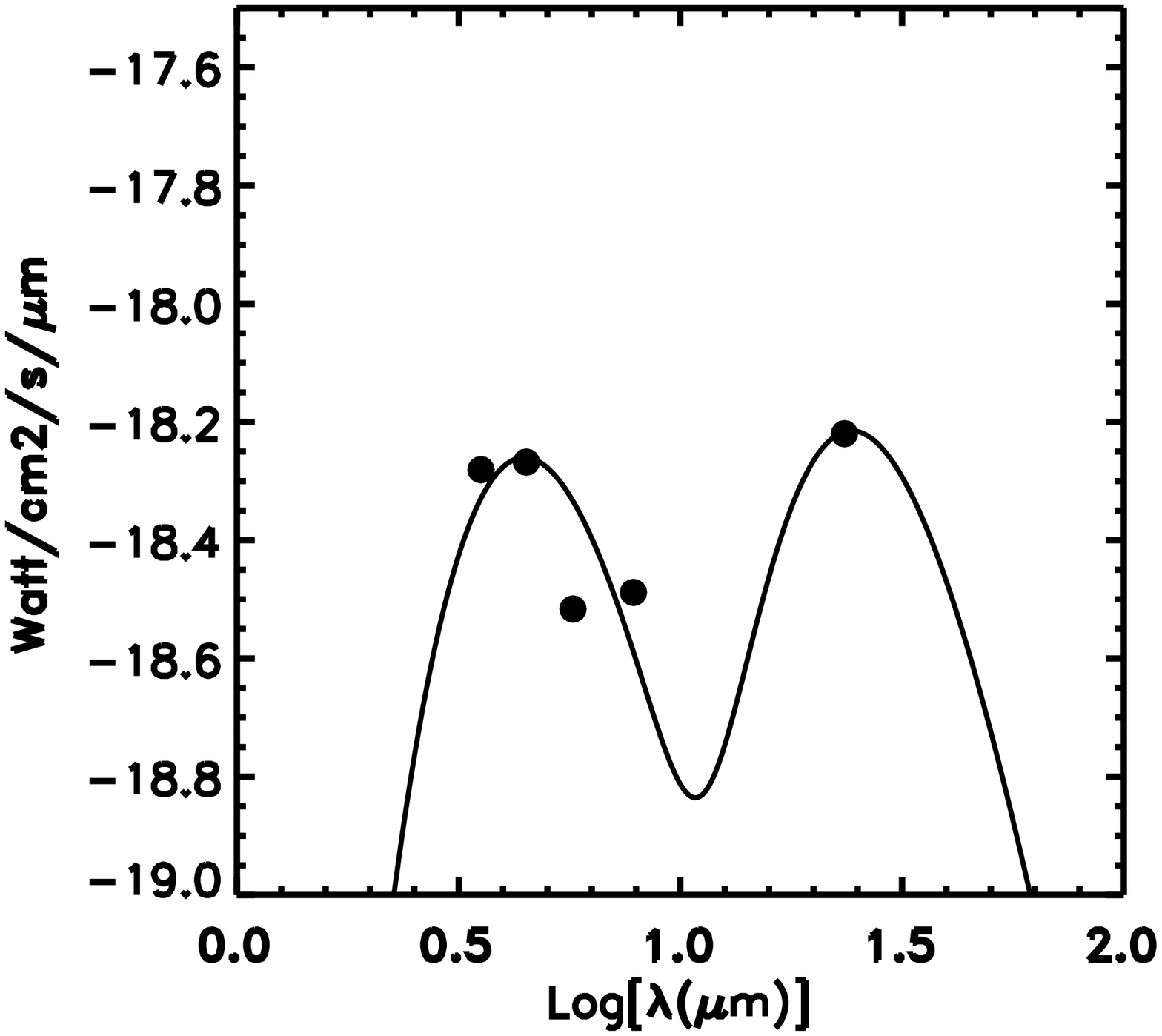}
   \end{center}
   \caption{MIR and FIR energy distributions measured by \emph{IRAS} in 1983 (left panel), \emph{MSX} in 1996-7 during the main obscuration event (middle panel) and \emph{Spitzer} in 2005-6 (right panel). Blackbody fits of 200K/60K (\emph{IRAS}), 300K (\emph{MSX}) and 550K/110K (\emph{Spitzer}) are also shown.}
   \label{fig:dust}
\end{figure*}

Longward of the $K_s$-band there is a strong excess which we attribute to thermal emission by dust grains. We followed the dust fitting procedures common when handling
dusty WCL Pop-I WR stars in the use of Colangeli et al. (1995) laboratory
measurements of the wavelength dependence of amorphous carbon dust (e.g. Williams et al. 2001). Our results are shown in Fig. \ref{fig:dust} for coherent combinations of the available photometry as described below. 

In the MIR there are two fortuitous combinations of photometry from two
independent missions. \emph{MSX} observed M~2-29 in 1996--7 during the obscuration event, while
\emph{Spitzer} IRAC reobserved it in 2005-6 at light maximum. M~2-29 is not listed
in the point source catalogues of \emph{MSX} (PSC2.3) nor in the GLIMPSE-II
source lists of \emph{Spitzer}. Therefore, we measured M~2-29 in the absolutely
calibrated \emph{MSX} images in all four long wavelength bands (8.3--21.3 $\mu$m).
Likewise we obtained photometry in the four IRAC bands (3.6--8.0 $\mu$m) to
which we applied the appropriate aperture corrections, and the MIPS 24 $\mu$m
band whose diffuse absolute calibration is equivalent to that of \emph{MSX} (Cohen 2009). 
Fitting blackbody distributions to the data we find in the obscuration event 300K grains match all the \emph{MSX} points, while outside of this both warm (550K) and cool (100K) grains are required to match the double peaked SED.

The earliest relevant FIR data are those by \emph{IRAS} taken during 1983. 
Two dust components can be seen but from very different temperatures than those observed by \emph{Spitzer}. 
\emph{IRAS} requires 200 K and 60 K grains to fit the 12--60 $\mu$m excess emission
and, of course, it cannot observe the warm components seen by \emph{MSX} and \emph{Spitzer} IRAC. 
The \emph{IRAS} 25 $\mu$m detection of 2.0 Jy (based on observations over a 6-month span) makes an
interesting comparison with both \emph{Spitzer} MIPS 24 $\mu$m in 2005--6 (1.15 Jy) 
and \emph{AKARI}'s All-Sky Survey 19 $\mu$m flux of 1.18 Jy during 2006--7. 
All three FIR measurements were made at light maximum. 
The recent \emph{AKARI} and \emph{Spitzer} data are in excellent agreement and 
confirm a real change since \emph{IRAS}, whose 
bright 25 $\mu$m observation has an uncertainty of $\pm$11\%. 
Note also that the temperature of the cool dust also cooled during this period.
These changes are clearly consistent with the high mass-loss rate of an evaporating dust disk.

\section{Similarities with yellow symbiotic stars}
\label{sec:sy}
As there is currently no evidence for an evolved companion to the Of(H) CSPN the chance that M~2-29 is a symbiotic star is remote. However, it is still instructive to draw this comparison to demonstrate that there is \emph{indirect} evidence for binarity in M~2-29. 
The closest sub-group of symbiotic stars to M~2-29 are the yellow or D'-type symbiotics (Schmid \& Nussbaumer 1993). Examples include AS~201 (Kohoutek 1987), V471~Per, StH$\alpha$~190 and HD~330036 (Schmid \& Nussbaumer 1993). Instead of the usual M-type giant or Mira companion of D-types, D'-types have yellow G-type giants (Belczy{\'n}ski et al. 2000). They also exhibit much cooler dust emission (200--400 K) than D-types (800--1000 K), a dense nebula of $N_e\sim10^6$--10$^7$ cm$^{-3}$, and a relatively cool WD of $T_\mathrm{eff}\sim50$--60 kK. In this work we have demonstrated M~2-29 shares all these properties except for the yellow giant companion. 

The evolutionary status of D' symbiotics is discussed by Jorissen et al. (2005). Their WD components are thought to have recently evolved from the AGB which explains the remnant dust emission, relatively cool WD temperatures and ionised nebulae (Schwarz 1991; Van Winckel et al. 1994). AS~201 in particular matches the nebula configuration of M~2-29 with a dense inner nebula surrounded by an older \emph{planetary} nebula thought to be ejected by the WD. Interaction with the wind of the cool giant companion creates the dense inner nebula. There is a small possibility this is happening in M~2-29 if an obscured yellow sub-giant is present. A sub-giant would not yet be luminous enough to rival the luminosity of the Of(H) CSPN ($M_V=-0.9$ mag at 7.4 kpc), especially if there is some additional obscuration from the dust. This would explain the its absence in our VLT FLAMES spectroscopy. 
Dusty cocoons appear in the symbiotic stars RX Pup (Miko{\l}ajewska et al. 1999), other symbiotic Miras (Gromadzki et al. 2009), the S-type yellow symbiotic AE Cir (Mennickent et al. 2008), and even in some single C-rich Miras like R Vol (Whitelock et al. 1997). All of these share with M~2-29 similar depths and durations of fading events due to dust. 
RX Pup is a particular interesting example where during obscuration all photospheric features of the Mira disappear (TiO bands in the optical and CO bands in the NIR), although pulsations can still be seen in the NIR. 

There are other explanations for dense inner nebulae in PNe but they all involve binary CSPN. Rodr\'iguez et al. (2001) found a dense circumstellar nebula in Hen 2-428, a bipolar PN which has since been found to have a double-degenerate close binary CSPN (Santander-Garc\'ia et al. 2010). We discount this possibility for M~2-29 as we found no evidence for a close binary (Sect. \ref{sec:small}). A more feasible scenario may be a very wide main-sequence companion as in EGB~6-like CSPN (Frew \& Parker 2010; Miszalski et al. 2010a). The archetype of this class exhibits a dense circumstellar nebula around a resolved late-M main-sequence companion \emph{and} cool dust at $T$$\sim$300 K despite the highly evolved nature of the CSPN (Liebert et al. 1989; Fulbright \& Liebert 1993; Bond 2009; Su et al. 2010). If this dust did not survive from the AGB phase, then the only explanation for its presence would be ongoing interactions with the companion. This may involve a mismatched interacting wind (Kenyon et al. 1993; Nussbaumer 2000; Miszalski et al. 2010a), however this is highly uncertain given the limited understanding of wide binary interactions in general. It is interesting to speculate that EGB~6-like CSPN could be the final evolutionary stage of M~2-29.

One way to unambiguously identify symbiotic stars is the detection of the Raman scattered O VI lines at $\lambda$6825 and $\lambda$7082 (Schmid 1989). Unfortunately, the low $T_\mathrm{eff}=50$--60 kK of M~2-29 and D'-type symbiotics (Schmid \& Nussbaumer 1993) is not hot enough to produce O VI which is only seen in the hottest symbiotics that have enough scattering particles available (see Miko{\l}ajewska et al. 1997). The Raman feature is therefore a sufficient but not necessary condition for a symbiotic star classification. Furthermore, even if M~2-29 had a primary hot enough to excite O VI, the curious appearance of the Raman lines in AE Cir \emph{only} during obscuration (Mennickent et al. 2008) means we might not expect to see them at all during light maximum when our VLT FLAMES spectra were taken. Another interesting example is CI Cygni which never shows the Raman lines and only episodically shows [Fe VII] lines (Miko{\l}ajewska \& Ivison 2001). 
Long-term spectroscopic monitoring of M~2-29 may similarly reveal evidence of a companion. 

\section{Conclusions}
\label{sec:conclusion}
We have studied the three PNe whose CSPN are known to show dust obscuration events: Hen~3-1333, NGC~2346 and M~2-29. Analysis of the lightcurves suggests a binary companion is not responsible for triggering the events but may be responsible for establishing the dust disks surrounding their CSPN. A particular focus was the unusual Galactic Bulge PN M~2-29 using the definitive OGLE lightcurve, deep medium-resolution VLT FLAMES mini-IFU spectroscopy and infra-red observations. A number of misconceptions surrounding M~2-29 were clarified during the course of the study. Our main conclusions are as follows:

\begin{itemize}
   \item An updated \cpds lightcurve was modelled for the first time using the simple RCB-star lightcurve model of GS92. A good fit was obtained for a dust outflow velocity of $v_\infty=225$ km s$^{-1}$ (De Marco \& Crowther 1998) and a dust formation radius of $97\pm11$ AU (Chesneau et al. 2006). The dust causing the obscuration events forms at the inner edge of the dust disk resolved by the VLTI. There is no possibility that a binary companion could trigger these events via interaction at periastron at such a large radius to produce the pseudo-periodic events in the lightcurve every $\sim5$ years. If PNe with dust obscuration events are the progeny of AGB stars with LPVs, then this argues against binarity as the cause for LSPs. 
   \item A similar application to the OGLE lightcurve of M~2-29 finds a plausible inner radius for a dust disk of $70\pm10$ AU. Mid-infrared spectroscopy shows the disk to be rich in crystalline silicates with no PAHs present. Recent \emph{AKARI} and \emph{Spitzer} observations at 24 $\mu$m confirm the flux has dropped from 2.0 Jy from \emph{IRAS} to 1.18 Jy reflecting the evaporation of the dust disk by the high mass-loss wind of the CSPN $\sim$10$^{-6}$ M$_\odot$ yr$^{-1}$. Stochastic variations in the wind and/or dust disk are likely to be responsible for the irregularity in obscuration events. 
   \item Analysis of the OGLE lightcurve finds no evidence for binarity as previously claimed by \hajduk. A colour change during the obscuration event is only caused by the dust and not a companion.
   \item A TMAP NLTE model was fitted to the VLT FLAMES spectrum of the CSPN to find $T_\mathrm{eff}=50\pm10$ kK and log $g=4.0\pm0.3$. We classified the CSPN as Of(H) following M\'endez et al. (1990) and M\'endez (1991). Many weak emission lines were found in the spectrum including Si~II, Si~III and Si~IV. These may not represent an overabundance of silicon, but rather some kind of interaction between the Of(H) wind and the silicate-rich dust disk. A gravity distance of $6.9\pm3.0$ kpc was found after assuming $M=0.60$ $M_\odot$ from evolutionary tracks (Miller Bertolami \& Althaus 2006). 
   \item \emph{HST} images of M~2-29 were used to independently derive the reddening to be $c=0.95\pm0.05$ or $E(B-V)=0.65$ mag. This value is free from the strong stellar H$\alpha$ contamination that effects previous ground based measurements and was used in calculating a statistical distance of $7.7\pm2.3$ kpc using the H$\alpha$ surface brightness-radius relation of Frew \& Parker (2006). A weighted average distance of $7.4\pm1.8$ kpc implies an ionised nebula mass of $0.08\pm0.05$ $M_\odot$ that is more consistent with PNe than symbiotic nebulae.
   \item Chemical abundances of M~2-29 were calculated for a spectrum of the outer nebula uncontaminated by the dense circumstellar nebula that contaminated most previous studies. The abundances are in agreement with TP97 and the oxygen abundance of 8.3 dex, together with the $7.4$ kpc distance and its Galactic coordinates (PN G004.0$-$03.0), clearly disproves the common misconception that M~2-29 is a Halo PN. \emph{M~2-29 is a Galactic Bulge PN}. 
   \item M~2-29 shares a number of similarities with D'-type symbiotic stars including a dense circumstellar nebula, an outer \emph{planetary} nebula, a cool $T_\mathrm{eff}=50$ kK CSPN and cool dust (100--300 K). We can rule out a yellow giant companion but a subgiant luminosity class or fainter cannot be ruled out as it may be obscured by dust or hidden in the glare of the Of(H) $M_V=-0.9$ CSPN. Further observations to search for such a companion are required for a definitive answer. 
\end{itemize}

\begin{acknowledgements}
   This work was conducted as part of the PhD thesis of BM at the Observatoire de Strasbourg and Macquarie University with additional valuable support from the framework of the European Associated Laboratory ``Astrophysics Poland-France''. BM thanks the CAMK staff for their hospitality and Observatoire de Strasbourg for additional travel support. JM is supported in part by the Polish Research Grant No. N203 395534 and TR is supported by the German Aerospace Center (DLR) under grant 05\ OR\ 0806. BM and AA thank Claudio Melo for his excellent support during our stay at Paranal and the provision of dark frames that were invaluable for data reduction. Thanks also to ESO and PICS/Observatoire de Strasbourg for travel support to Chile. The OGLE project has received funding from the European Research Council under the European Community's Seventh Framework Programme (FP7/2007-2013) / ERC grant agreement no. 246678 to AU.
\end{acknowledgements}

\end{document}